\documentclass{article}

\usepackage{microtype}
\usepackage{booktabs} 
\usepackage{subfigure}
\usepackage{graphicx}
\usepackage{pifont}
\usepackage{enumitem}
\usepackage{tikz}
\usepackage{amsmath}

\usepackage{filecontents}

\usepackage{soul}
\usepackage{xspace} 
\usepackage{url}
\usepackage{multirow}

\usepackage[skins]{tcolorbox}
\usepackage{array}

\newcommand{\skmsg}{{SKMSG}\xspace}

\newcommand{\name}{LIFL\xspace}
\newcommand{\flame}{Flame\xspace}

\newcommand{\ie}{{\em i.e., \/}}
\newcommand{\eg}{{\em e.g., \/}}
\newcommand{\etc}{{\em etc. \/}}

\newcommand{\hide}[1] {}

\newcommand{\knote}[1]{{\color{red}[KK: #1]}}
\newcommand{\sxnote}[1]{{\color{brown} [SX: #1]}}
\newcommand{\mlnote}[1]{{\color{orange} [ML: #1]}}

\definecolor{ao(english)}{rgb}{0.0, 0.5, 0.0}
\newcommand{\cut}[1]{{\color{red} CUT: [#1]}}

\newcommand{\secref}[1]{\S\ref{#1}}

\newcommand{\textiu}[1]{\noindent\textit{\underline{#1}}}
\newcommand{\textib}[1]{\noindent\textbf{\textit{#1}}}
\newcommand{\textbt}[1]{\textbf{\texttt{#1}}}

\definecolor{blue(munsell)}{rgb}{0.0, 0.5, 0.69}
\definecolor{airforceblue}{rgb}{0.36, 0.54, 0.66}
\definecolor{lincolngreen}{rgb}{0.11, 0.35, 0.02}
\definecolor{lasallegreen}{rgb}{0.03, 0.47, 0.19}
\definecolor{amber(sae/ece)}{rgb}{1.0, 0.49, 0.0}
\definecolor{majorelleblue}{rgb}{0.38, 0.31, 0.86}
\newcommand{\new}[1]{{\color{blue}#1}}

\setlist[itemize]{leftmargin=0.4cm}

\newcommand{\circlednumber}[1]{\textcircled{\footnotesize \textbt{#1}}}
\newcommand{\X}{$\times$\xspace}

\usepackage{hyperref}
\makeatletter
\def\UrlAlphabet{%
      \do\a\do\b\do\c\do\d\do\e\do\f\do\g\do\h\do\i\do\j%
      \do\k\do\l\do\m\do\n\do\o\do\p\do\q\do\r\do\s\do\t%
      \do\u\do\v\do\w\do\x\do\y\do\z\do\A\do\B\do\C\do\D%
      \do\E\do\F\do\G\do\H\do\I\do\J\do\K\do\L\do\M\do\N%
      \do\O\do\P\do\Q\do\R\do\S\do\T\do\U\do\V\do\W\do\X%
      \do\Y\do\Z}
\def\UrlDigits{\do\1\do\2\do\3\do\4\do\5\do\6\do\7\do\8\do\9\do\0}
\g@addto@macro{\UrlBreaks}{\UrlOrds}
\g@addto@macro{\UrlBreaks}{\UrlAlphabet}
\g@addto@macro{\UrlBreaks}{\UrlDigits}
\makeatother

\usepackage[accepted]{mlsys2023}

\mlsystitlerunning{\name: A Lightweight, Event-driven Serverless Platform for Federated Learning}

\begin{document}

\twocolumn[
\mlsystitle{\name: A Lightweight, Event-driven Serverless Platform for Federated Learning}



\mlsyssetsymbol{equal}{*}

\begin{mlsysauthorlist}
\mlsysauthor{Shixiong Qi}{aff1}
\mlsysauthor{K. K. Ramakrishnan}{aff1}
\mlsysauthor{Myungjin Lee}{aff2}
\end{mlsysauthorlist}

\mlsysaffiliation{aff1}{University of California, Riverside}
\mlsysaffiliation{aff2}{Cisco Research}

\mlsyscorrespondingauthor{Shixiong Qi}{sqi009@ucr.edu}
\mlsyscorrespondingauthor{K. K. Ramakrishnan}{kk@cs.ucr.edu}
\mlsyscorrespondingauthor{Myungjin Lee}{myungjle@cisco.com}

\mlsyskeywords{Federated Learning, Serverless Computing, eBPF}

\vskip 0.3in

\begin{abstract}
Federated Learning (FL) typically involves a large-scale, distributed system with individual user devices/servers training models locally and then aggregating their model updates on a trusted central server. 
Existing systems for FL often use an always-on server for model aggregation, which can be inefficient in terms of resource utilization. They may also be inelastic in their resource management. This is particularly exacerbated when aggregating model updates at scale in a highly dynamic environment with varying numbers of heterogeneous user devices/servers. 

We present \name, a lightweight and elastic serverless cloud platform with fine-grained resource management for efficient FL aggregation at scale. \name is enhanced by a streamlined, event-driven serverless design that eliminates the individual heavy-weight message broker and replaces inefficient container-based sidecars with lightweight eBPF-based proxies. We leverage shared memory processing to achieve high-performance communication for hierarchical aggregation, which is commonly adopted to speed up FL aggregation at scale. We further introduce locality-aware placement in \name to maximize the benefits of shared memory processing. \name precisely scales and carefully reuses the resources for hierarchical aggregation to achieve the highest degree of parallelism while minimizing the aggregation time and resource consumption. Our experimental results show that \name achieves significant improvement in resource efficiency and aggregation speed for supporting FL at scale, compared to existing serverful and serverless FL systems.

\end{abstract}
]



\printAffiliationsAndNotice{}  




\section{Introduction}\vspace{-2mm}

Federated Learning (FL~\cite{fedavg}) enables collaborative model training across a network of decentralized devices/machines while keeping individual user data secure and private. In FL, instead of sending raw data to a central server, models are trained on individual devices/machines using local data, and only the model updates are shared and aggregated to create a global model.

To support FL at scale, {\it hierarchical aggregation} is often adopted to increase the service capacity for model aggregation~\cite{google-fl-mlsys19, adafed}. This can accommodate a large number of clients and handle a substantial volume of model updates, avoiding potential slow-down of the aggregation process. In the process, each level performs intermediate aggregation, combining the updates from lower-level aggregators or clients. 

Existing FL frameworks (\eg Google's FL stack~\cite{google-fl-mlsys19}, Meta's PAPAYA~\cite{papaya-mlsys2022}) adopt a static, always-on\footnote{meaning that aggregators are up all the time within a round.} deployment to support model aggregation.
However, in a dynamic FL environment, it's difficult to have a one-size-fits-all service capacity for model aggregation. 
System heterogeneity (\ie different hardware capabilities) and a dynamically varying number of participating clients in each round
require frequent adjustments of the capacity so that the aggregation service 
effectively uses resources on demand and avoids significant resource wastage. 

Serverless computing promises to provide an event-driven, resource-efficient cloud computing environment, enabling services to use resources on demand~\cite{serverless}.
Running FL model aggregation service as serverless functions can right-size the provisioned resources and reduce resource waste compared to an always-on aggregation server implementation. In addition, stateless processing by serverless functions makes it easy to support continual updates to the aggregation hierarchy. 
By increasing the capacity of aggregation through a hierarchy of serverless aggregators, model aggregation in FL can be executed in parallel, responding to  increasing loads from trainer model updates.

However, the excessive overhead in current serverless frameworks, caused by the loose coupling of data plane components~\cite{spright-sigcomm22}, is a barrier to achieving efficient and timely aggregation, compared to a monolithic serverful design. Further, the use of individual, constantly-running components (e.g., container-based sidecars) in current serverless frameworks is inefficient and sacrifices much of the benefit of serverless computing. 
This prompts us to create a more streamlined, responsive serverless framework that is tailored to achieve just-in-time FL aggregation on demand.

We introduce \name, a lightweight serverless platform for FL that uses hierarchical aggregation to achieve parallelism in aggregation and exploits intra-node shared memory processing to reduce data plane overheads. 
\name also utilizes a locality-aware placement policy to maximize the benefits of the intra-node shared memory data plane. Unlike typical serverless platforms that use a heavyweight sidecar implemented as a separate container, \name seeks to eliminate this wasteful overhead by taking advantage of eBPF-based event-driven processing. This ensures that resource usage is truly load-proportional. 
Instead of depending on inaccurate, threshold-based autoscaling, \name uses hierarchy-aware autoscaling to precisely adjust the capacity of model aggregation
to match the incoming load.
We also use a policy of reusing runtimes
to sidestep the impact of startup delay on the model convergence time, while also improving resource efficiency of aggregation. 
\name favors eager aggregation to enable timely aggregation,  reducing the queuing time for model updates. By harnessing the capabilities of \name, FL systems can achieve efficient resource utilization and reduced aggregation time. 
\name is available at~\cite{flame-cisco}.

We highlight the contributions of \name below:

\noindent\textbf{(1)} 
\name's enhanced data plane achieves 3\X (compared to serverful) and 5.8\X (compared to serverless) latency reduction on transferring a relatively heavyweight ResNet-152 model update within the aggregation hierarchy (intra-node).

\noindent\textbf{(2)} 
\name's locality-aware placement can maximize shared memory processing, achieving up to 2.1\X additional latency reduction on aggregating a batch of updates in a round (details in \S\ref{sec:evaluation}).
After applying hierarchy-planning, aggregator reuse, and eager aggregation, \name can further obtain 1.5\X latency reduction.
The enhanced orchestration also helps improve efficiency, saving up to 2\X CPU consumption compared to simply using the enhanced data plane.

\noindent\textbf{(3)}
Our evaluation with a real FL workload using ResNet-18 and 120 simultaneous active clients (the total number of clients used is 2,800) shows that the combination of \name's enhanced data and control planes achieve \textbf{5\X} and \textbf{1.8\X} less CPU cost and reduces \textbf{2.7\X} and \textbf{1.6\X} on time-to-accuracy (70\% accuracy level), compared to existing serverless and even serverful FL systems. We also train a relatively heavyweight ResNet-152 model. \name spends 1.68\X less time to reach 70\% accuracy than existing serverless FL systems, while using 4.23\X fewer CPU cycles.

    \begin{figure}[t]
    \centering
        \includegraphics[width=\columnwidth]{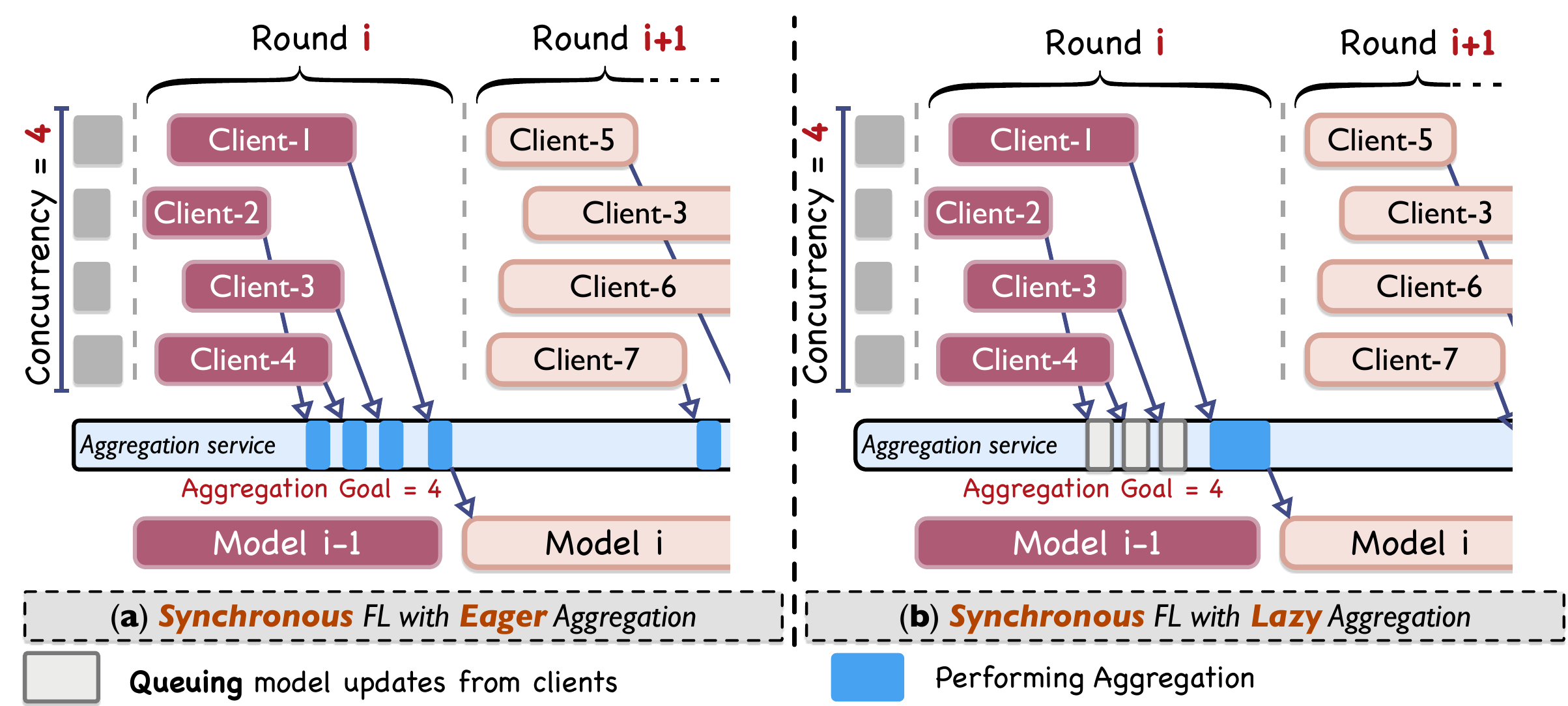}\vspace{-7mm}
    \caption{Synchronous FL with different aggregation timing (``Eager'' and ``Lazy'')~\cite{google-fl-mlsys19, jit-agg}.} 
    \vspace{-5mm}
    \label{fig:sync-async-fl}
    \end{figure}

\vspace{-2mm}\section{Background and Challenges}\label{sec:background}\vspace{-2mm}

\subsection{Basics of Federated Learning}\label{sec:sync-async-fl}\vspace{-2mm}

\hide{\noindent\textbf{Synchronous FL and Asynchronous FL:}
FL protocols can be broadly classified into {\it synchronous} or {\it asynchronous} FL.
{\it Synchronous} FL enforces synchronization among clients, ensuring that
all selected clients use the same version of global model for local training.
It provides a clear separation between consecutive rounds, based on the evolution of the model version (Fig.~\ref{fig:sync-async-fl} (a) and (c)). 
In contrast, {\it asynchronous} FL does not separate the aggregation process into different rounds. It allows clients to train on a previous (\ie stale) global model (Fig.~\ref{fig:sync-async-fl} (b) and (d)).
Asynchronous FL offers greater flexibility in terms of timing and frequency of model updates and allows for higher client utilization~\cite{pisces-socc22, papaya-mlsys2022}.}

\noindent\textbf{FL aggregation:} 
Aggregation in FL is a process of building a global model from individually trained model updates. The {\it aggregation goal}, $n$  specifies the expected number of model updates to be received before the global model is updated to a new version. Thus, it dictates the number of selected clients for training. The aggregation process is abstracted as:

\vspace{-6mm}
\begin{equation}
w_{i} = f(\{ (w^k_{i}, \mathcal{A}^k_{i})~|~1 \le k \le n \}).
\end{equation}
\vspace{-8mm}

Here $f(\cdot)$ is an aggregation function, $w^k_{i}$ is $k$-th local model update for global model version $i$, and $\mathcal{A}^k_{i}$ is auxiliary information for aggregation. For the {\it FedAvg} algorithm~\cite{fedavg}, $f(\cdot) = \sum^n_{k=1}w^k_i c^k_i / T_i$. $T_i = \sum^n_{k=1} c^k_i$ and $\mathcal{A}^k_{i}$ is  $c^k_i$ (the number of data samples).



\vspace{-2mm}\noindent\textbf{Eager aggregation and Lazy aggregation:}
Based on the 
timing to trigger the aggregation, we can classify the model aggregation to be ``\textit{eager}'' or ``\textit{lazy}''~\cite{jit-agg}:
{\it Eager} aggregation allows aggregation to happen whenever an update is received, leading to more  flexible and dynamic timing of the aggregation process. {\it Lazy} aggregation operates on a delayed schedule, where model updates that arrive early are queued without being aggregated immediately.
\hide{{\it Lazy} aggregation is always feasible. However, {\it eager} aggregation is possible only if an aggregation function doesn't rely on the information constructed from other clients. For instance, FedAvg~\cite{fedavg} adopts an aggregation function that is hard to apply eager aggregation as it uses the number of data samples from all participating clients to compute each client's contribution as its weight. To allow {\it eager} aggregation, we consider a relaxed version of FedAvg that only uses the number of clients instead of their data sample sizes. In contrast, FedBuff~\cite{fedbuff}, an algorithm for asynchronous FL, relies only on the staleness of a model update from a client. Thus, {\it eager} aggregation is allowed.} 
Fig.~\ref{fig:sync-async-fl} shows the two different aggregation methods 
for synchronous FL. For instance, the \textit{eager} method is feasible for FedAvg with cumulative averaging. 

\hide{\noindent\textbf{FL-specific hyperparameters:}
There are several FL-specific hyperparameters that affect the behavior of the FL process, including {\it concurrency} and {\it aggregation goal}. {\it Concurrency} (Fig.~\ref{fig:sync-async-fl}), in both synchronous and asynchronous FL, indicates the number of participating clients, performing local training in parallel. A certain number of clients are selected to satisfy {\it concurrency}. 
\knote{main point}The {\it aggregation goal} specifies the expected number of model updates to be received before the global model is updated to the new version. This determines the timing for the model to be updated to the new version.}

\hide{\mlnote{Shixiong's plan is to use the details in the following two paragraphs in sec 6.4. Otherwise, we can remove them.}

In Synchronous FL, over-provisioning the number of clients (over-selection) is often used to mitigate the impact of stragglers, specifying an {\it aggregation goal} to be less than the training concurrency. This leads to wasted effort by slow clients, whose updates may arrive late, thus discouraging participation by slow clients~\cite{refl}. Their speed may often be negatively correlated with the data quality~\cite{pisces-socc22}, ultimately affecting the model quality of Synchronous FL.

Asynchronous FL also requires that the {\it aggregation goal} be smaller than the concurrency, to avoid being impeded by potential stragglers. However, since Asynchronous FL allows stale updates to still be aggregated, {\it no} client effort will be wasted compared to the Synchronous FL. This ensures high client utilization throughout the FL process and potentially a more diverse dataset contributed by slow clients.}

\vspace{-2mm}\subsection{Anatomy of Systems for Federated Learning}\vspace{-2mm}
Designing a system to support FL at a \textbf{\textit{large scale}} is essential, as a larger 
number of participants means a more diverse and representative dataset. 
It improves the model's ability to capture complex patterns and unseen relationships in the data.
These benefits 
help the model generalize in real-world deployments, \eg Google's FL stack 
has been used to serve $\sim$$10M$ devices daily and 
$\sim$$10K$ devices participate in FL training simultaneously~\cite{google-fl-mlsys19}.

\begin{figure}[t]
  \centering
  \includegraphics[width=\columnwidth]{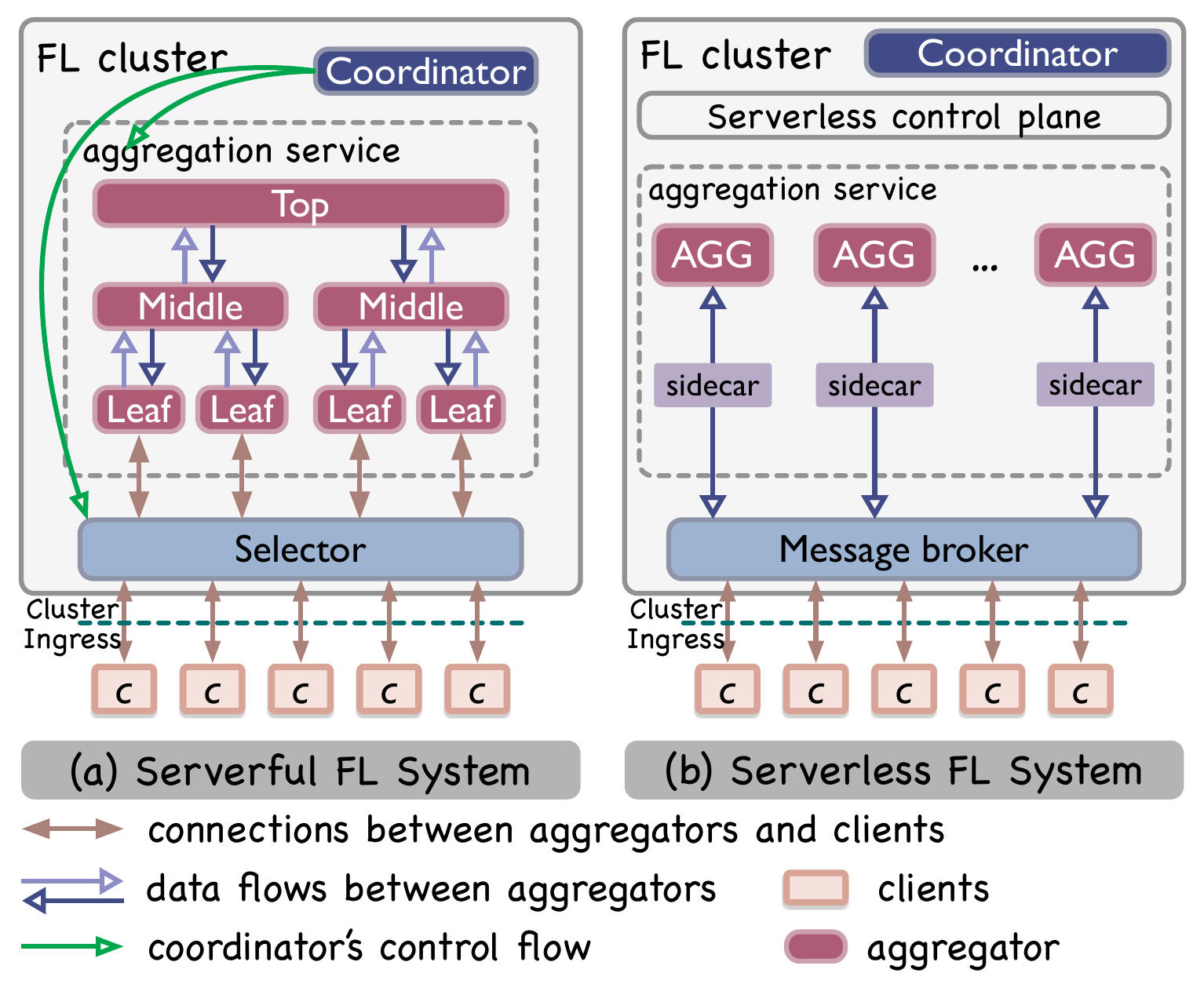}\vspace{-9mm}
  \caption{Generic architectures for FL systems: (a) Serverful FL systems~\cite{google-fl-mlsys19, papaya-mlsys2022}; (b) Serverless FL systems~\cite{lambda-fl, adafed, fedkeeper}. Note that for simplicity, we skip the hierarchy in the diagram (b).}
  \vspace{-7mm}
  \label{fig:fl-system}
\end{figure}

Fig.~\ref{fig:fl-system} depicts key architectural components that are needed to ensure the success of FL at scale.\footnote{We adopt the terminology of FL system components from \cite{google-fl-mlsys19} and \cite{papaya-mlsys2022}.}
These components work together to enable the collaborative and decentralized training process in FL. 
In addition to the {\bf aggregator} and the {\bf client},
the {\bf coordinator} oversees the flow of FL operations. It acts as an orchestrator that facilitates seamless interactions among aggregators, selectors, and clients by applying the client selection scheme and instructing the selector to map the selected clients to backend aggregators~\cite{google-fl-mlsys19}.
The {\bf selector} plays two roles. 
First, it ensures that a diverse set of clients participate in the FL process to capture a representative sample of the distributed data.
Second, 
it acts as a gateway that mediates communication (\ie queuing, load balancing) between (leaf) aggregators and clients~\cite{google-fl-mlsys19, papaya-mlsys2022}.
\hide{The {\bf aggregator} combines the local model updates into a global model that represents the collective knowledge learned from potentially widely distributed data, using suitable aggregation algorithms (\eg FedAvg~\cite{fedavg}, FedBuff~\cite{fedbuff}).
The {\bf clients} actively train their models on their own dataset and contribute model updates to an aggregator.
This approach simplifies orchestration in FL systems, especially at large scale, by clearly separating the role and functionality of different components. }


\hide{
\noindent\textbf{Coordinator:}
The coordinator plays a pivotal role in orchestrating the  overall flow of FL operations. It facilitates seamless interactions among the aggregators, selectors, and clients. The coordinator is responsible for applying the client selection scheme and directing the selector to determine which clients participate in the FL process. 
To aggregate the incoming model updates, the coordinator spawns aggregators.
Additionally, it instructs the selector to redirect the selected clients to different aggregators, ensuring a balanced distribution of the aggregation tasks.
}

\hide{\noindent\textbf{Aggregator:}
The aggregator subsystem is responsible for aggregating the model updates received from participating clients. It combines the local model updates into a global model that represents the collective knowledge learned from the distributed data, using appropriate aggregation algorithms (\eg FedAvg~\cite{fedavg}, FedBuff~\cite{fedbuff}).}

\hide{
\noindent\textbf{Selector:}
The selector plays a hybrid role in both the {\it control} and {\it data} planes in the FL system. As a control plane component, the selector ensures that a diverse set of clients participate in the FL process to capture a representative sample of the distributed data.
It determines which clients will be included in each training round based on various criteria received from the coordinator, such as device availability, data quality~\cite{oort, refl, pisces-socc22}, \etc. 
As a data plane component, the selector acts as a gateway that mediates communication between (leaf) aggregators and clients, helping to balance the load across leaf aggregators. The selector is also able to buffer the model updates from clients, when the aggregators are being spawned by the coordinator~\cite{google-fl-mlsys19, papaya-mlsys2022}.
}

\hide{\noindent\textbf{Client (Trainer):}
Clients are the distributed devices or nodes that participate in FL by training local models on their own dataset. During each training round, clients train their models using their local data and contribute model updates to the aggregator. The client also receives global model updates from the aggregator to further refine its local model.}

\noindent\textbf{Need for hierarchical aggregation:}
The growing number of participating clients in FL requires the system to be scalable to accommodate the computational requirements of aggregating model updates from a large number of distributed clients.
This primarily motivates the use of {\it hierarchical aggregation} potentially involving multiple levels of aggregation in the FL process~\cite{google-fl-mlsys19, adafed}, as depicted in Fig.~\ref{fig:fl-system} (a).
Essentially, hierarchical aggregation is structured as a single-rooted tree.
Each level in the tree includes multiple parallel aggregation tasks that are executed by one of potentially multiple aggregators.
The communication during the hierarchical aggregation task takes place across multiple levels: 
The model updates from smaller subgroups of clients are aggregated by the lower-level aggregators (\ie leaf) and passed onto higher-level aggregators (\ie top),
until a global model is obtained. 
This parallel aggregation at the lower levels can provide speedup and reduce queueing of model updates. 

\hide{Hierarchical aggregation can further mitigate the negative impact (\eg heavyweight synchronization between aggregators/clients, injecting noise to model updates that can impact model accuracy) of operating secure aggregation~\cite{secure-aggregation-ccs17} and differential privacy~\cite{varun-hdp} at scale, which are necessary enhancements to ensure client privacy in FL.}

\hide{
\sxnote{Data privacy can be shortened. It seems to be less relevant in this paper.}
\noindent\textbf{Data Privacy Enhancement:}
Though FL protects users' privacy as data remains on local devices/servers, eliminating the need to transfer raw data to a centralized server, there still is a risk of exposing sensitive information about individual users~\cite{}. 
By inspecting intermediary model updates from individual clients, private training data can be reconstructed. 

To preserve privacy in the face of local model update exposure, FL training is often applied with data privacy enhancement, such as secure aggregation~\cite{} and differential privacy~\cite{}, to mitigate privacy risks caused by the exposure of local model updates.
Secure aggregation is essentially a variant of Multi-party Computation (MPC) which ensures that the central aggregator only has an aggregated view of model updates without being able to inspect the contribution of individual clients. 
Differential privacy, on the other hand, protects the privacy by adding carefully calibrated noise to local model updates before being aggregated. This hides the contribution of any specific user's data while still preserving the statistical properties required for accurate model training.



However, the use of this privacy enhancement is a double-edged sword: the secure aggregation has to run a heavyweight MPC protocol to negotiate zero-sum masks between individual clients on a per-round basis, which imposes an excessive communication overhead and delayed aggregation times~\cite{google-fl-mlsys19, adafed}. Differential privacy, without careful calibration of the added noise, may degrade the model quality. These negative sides of the privacy enhancement for FL could be further amplified when working with a system for FL at scale, especially when dealing with a large number of clients.

\sxnote{One suggestion is to move this paragraph to the end of the ``Hierarchical aggregation'' paragraph and remove the details of enhancement of privacy preserving.}
The use of hierarchical aggregation can alleviate the side effects of data privacy enhancements {\it at scale}, \eg reducing the communication cost of secure aggregation~\cite{google-fl-mlsys19}, balancing privacy-utility trade-off when using differential privacy for FL~\cite{varun-hdp}. 





}

\vspace{-2mm}\subsection{Motivation and Challenges for Serverless FL}\label{sec:shifting}\vspace{-2mm}

State-of-the-art FL systems~\cite{google-fl-mlsys19, papaya-mlsys2022} rely on a ``serverful'' design 
that relies on a fixed pool of dedicated resources (\eg CPU and memory), using a pool of provisioned VMs. 
Resizing the pool often takes a 
long time (e.g., 6 to 45 minutes on AWS~\cite{aws-resize}).
\hide{We demonstrate the penalties of serverful systems that do not dynamically scale the aggregation service to match the demand, in \S\ref{sec:system-eval}. This results in either poor responsiveness or resource wastage.}
Serverless computing, on the other hand, brings fine-grained resource elasticity
by provisioning functions (typically as containers) dynamically based on demand, ensuring that the right amount of resources is allocated only when needed. 

In FL, serverless computing can be used to provide efficient model aggregation, adapting to varying numbers of 
clients.
It eliminates the need to maintain dedicated resource pools for the aggregation service, thereby improving overall efficiency compared to the current ``serverful'' deployments. 


%


\noindent\textbf{Prior Work on Serverless FL.} 
A number of FL system designs have been proposed using serverless computing~\cite{lambda-fl, adafed, fedless}. 
A common abstract architecture of a serverless FL system and its key components is shown in Fig.~\ref{fig:fl-system} (b).
But, prior approaches still face the following challenges:

\textiu{Indirect networking:}
Unlike a ``serverful'' design (Fig.~\ref{fig:fl-system} (a)), a serverless FL system executes aggregators as serverless functions. Serverless function chaining can support hierarchical aggregation as well as communication between aggregators. However, because serverless functions are ephemeral and stateless (and thus unable to retain stateful information like routes), these chains typically only support {\it indirect} networking between functions. 
This raises the need for a stateful, persistent networking component (Fig.~\ref{fig:fl-system} (b)), such as a message broker or external storage services,\footnote{For consistency, we use the generic term ``message broker'' to denote such a networking component throughout this paper.} to maintain routes and exchange messages~\cite{spright-sigcomm22}.
However, having such a networking component in the internal datapath between serverless functions adds unnecessary overhead (20\% added delay as in Fig.~\ref{fig:hier-latency}).


\vspace{-2mm}\textiu{Inefficient message queuing:}
In addition to supporting function chaining, the message broker (Fig.~\ref{fig:fl-system} (b)) also acts as a message queue to buffer incoming model updates from clients while  aggregators are being spawned by the serverless control plane~\cite{lambda-fl, adafed}. 
However, the message broker and dedicated  queues add overhead and delay to the aggregation service. 

\vspace{-2mm}\textiu{Heavyweight sidecar:} 
Scheduling serverless functions typically requires metrics collection, often using a sidecar.
This container-based sidecar introduces additional network processing
in the datapath, requiring the interception and forwarding of model updates. 
This leads to complex data pipelines (involving extra communication hops between aggregators) and increased communication overheads due to the reliance on kernel-based networking~\cite{spright-sigcomm22}.

\vspace{-2mm}\textiu{Application-agnostic, simple, autoscaling:}
Current serverless autoscaler designs typically rely on a simplistic threshold based on user input (\eg request per second, concurrency) for scaling decisions~\cite{autoscaling-openfaas, autoscaling-knative}, often being unaware of application needs.
This design, agnostic of the hierarchical structure of FL aggregation, is limited in its ability to optimize the system to  maximize parallelism, \ie the number of levels and the number of aggregators at each level.
Looking at Fig.~\ref{fig:fl-system} (a), as we go up the levels in the hierarchy, fewer aggregators are needed. 
This can be leveraged to potentially reuse the lower-level aggregators as we proceed up the hierarchy. 
Further, since hierarchical aggregation uses function chaining, current ``reactive'' autoscaling designs lead to a cascading effect~\cite{graf-conext} of the cold-start delays when scaling a function chain.



\vspace{-2mm}\textiu{Locality-agnostic placement:}
{\it Intra-node} communication can be faster than {\it inter-node} communication by avoiding a lot of networking overheads~\cite{cni-tnsm} and using state-of-the-art serverless data plane designs with {\it shared memory}~\cite{faasm, spright-sigcomm22, pheromone-nsdi}. 
However, leveraging the benefits of shared memory effectively can be challenging when dealing with a large hierarchy of aggregators that cannot be accommodated within a single node. This requires careful function placement by taking into account the impact of communication between aggregators. Inter-node communication typically still uses kernel-based networking.


\hide{\noindent\textbf{Summary:}
\knote{I suggest we can't afford this paragraph. Hide it.}
As discussed, existing serverless designs face several challenges: heavyweight hierarchy data plane, 
costly message queuing, inefficient sidecar design, and suboptimal orchestration scheme in determining the hierarchy planning and aggregator placement.
Taken together, these limitations dampen the potential advantages of a serverless FL system, resulting in higher resource usage without fully realizing the expected benefits. They can also 
impede the timely convergence of ML models.}

\vspace{-2mm}\section{\name Overview}\label{sec:system-design}\vspace{-2mm}

We aim to address the 
aforementioned limitations (\secref{sec:shifting}) and develop \name---a high-performance, lightweight, and elastic serverless platform for FL,
utilizing hierarchical aggregation.
We focus on the following innovations of \name:

    \begin{figure}[t]
    \centering
        \includegraphics[width=\linewidth]{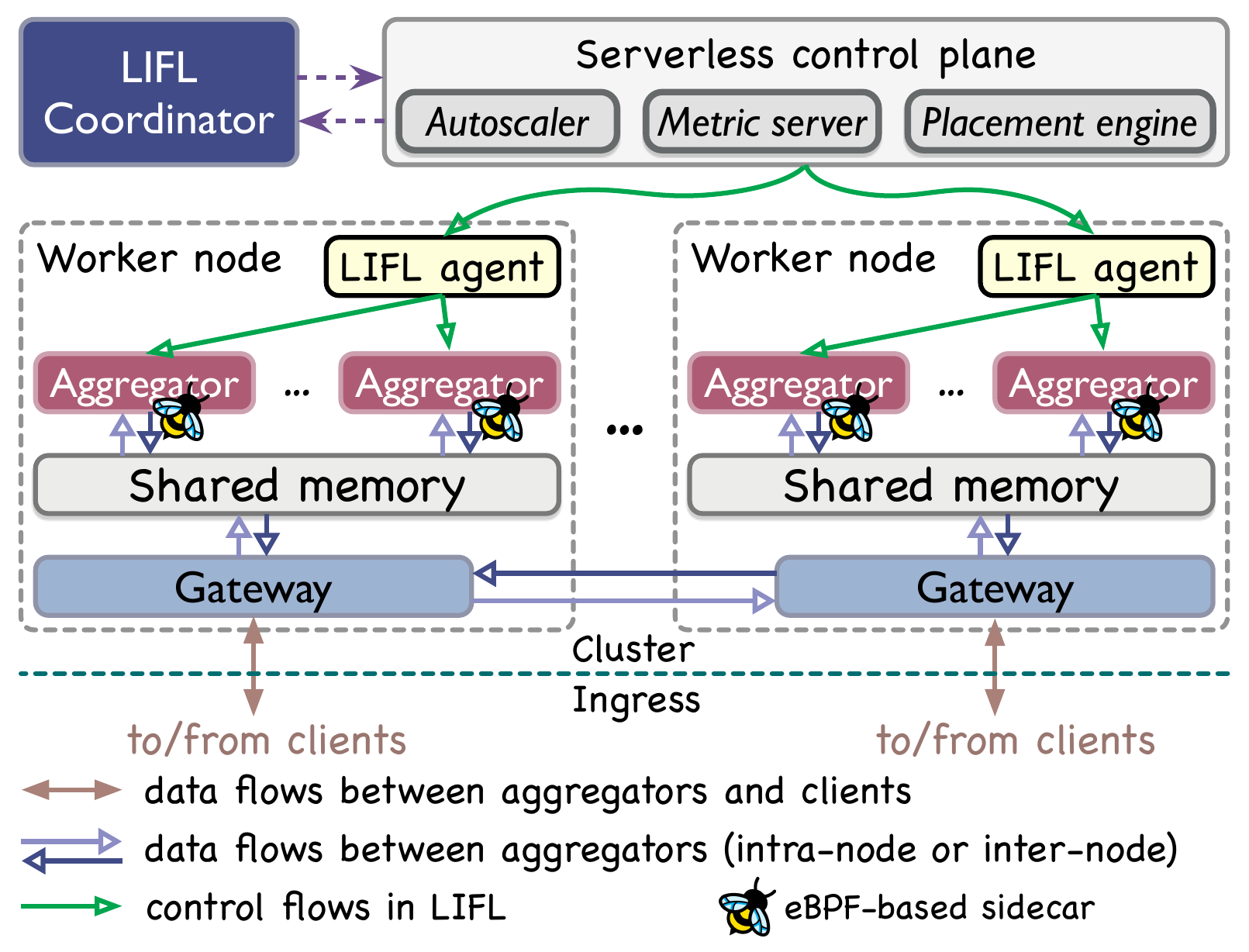}
        \vspace{-7mm}
    \caption{The overall architecture of \name.}
    \vspace{-4mm}
    \label{fig:system-overview}
    \end{figure}

    \vspace{-2mm}\noindent{\bf (1) High-performance intra-node dataplane:} 
    \name incorporates shared memory processing to provide a {\it zero-copy} communication channel between FL aggregators placed on the same node (\S\ref{sec:shm-support}). This avoids heavyweight kernel networking overheads, especially data copies~\cite{qizhe-host}, as model updates are often large, \eg a model update from 
    ResNet-152~\cite{resnet} is $\sim$230~MBytes.
    Shared memory can also eliminate other overheads such as protocol processing, serialization/de-serialization, kernel/userspace boundary crossing, and interrupts.
    
    \vspace{-2mm}\noindent{\bf (2) In-place message queuing:}
    We extensively leverage shared memory in \name to offer ``in-place'' message queuing (\S\ref{sec:msg-queue}). Messages (\ie model updates) from selected clients are directly buffered in shared memory and can be instantly accessed by the aggregators when they are ready. This eliminates dedicated message queues and their associated queuing delays.
    
    \vspace{-2mm}\noindent{\bf (3) Lightweight eBPF-based sidecar:} 
    We incorporate the extended Berkeley Packet Filter (eBPF~\cite{ebpf}) into \name to build a lightweight sidecar (\S\ref{sec:ebpf-sidecar}) to provide important functionality, \eg metrics collection. Unlike 
    a container-based sidecar,
    \name's sidecar runs as eBPF code attached at in-kernel hooks, avoiding the need for dedicated resources.
    We further utilize the eBPF-based sidecar to support {\it direct} networking between aggregators (\S\ref{sec:routing}), completely replacing the message broker. 
    
    \vspace{-2mm}\noindent{\bf (4) A cost-effective orchestration heuristic:}
    \name orchestrates the model aggregation to fully exploit the improved serverless dataplane by employing several strategies:
    (1) locality-aware placement that partitions levels with large traffic into node-affinity groups to make the best use of shared memory processing (\S\ref{sec:placement});
    (2) hierarchy-aware scaling that dynamically adjusts the configuration of hierarchical aggregation (\S\ref{sec:hier-plan}), and
    (3) opportunistic reuse of the aggregator runtime from a lower level (\S\ref{sec:reuse}).

\noindent\textbf{Architectural overview of \name:}
Fig.~\ref{fig:system-overview} shows the overall architecture of \name.
\name maintains a shared memory object store on each worker node to enable zero-copy communication between aggregators.
To support in-place message queuing, \name introduces a gateway on each worker node that receives model updates from remote clients. The gateway performs a consolidated, one-time payload processing to queue the received model updates to shared memory. 
Each aggregator in \name has attached to it an eBPF-based sidecar 
for lightweight metrics collection. 
Aggregators in \name are stateless, so new ones start without state synchronization upon an aggregator failure. \name detects client failures with keep-alive heartbeats and enhances resilience by over-provisioning the number of clients.
In the control plane, a \name agent is deployed on each worker node to manage the lifecycle (\eg creation, termination) of aggregators, following instructions from the \name control plane. 
The \name coordinator, a cluster-wide control plane component, is used for interactions between the FL job designer (ML engineer) and the serverless control plane (\eg autoscaler, placement engine).\footnote{Note: Even though the serverless control plane has serverful, always-on components (\eg autoscaler, placement engine), are shared and their overheads are amortized across multiple workloads, especially at scale.}
It works with the serverless control plane to execute \name's orchestration flow (\S\ref{sec:lifl-orches}).

\vspace{-2mm}\section{Optimizing the Serverless Data-Plane in \name}\label{sec:dp-design}\vspace{-2mm}

\subsection{Shared Memory for Hierarchical Aggregation}\label{sec:shm-support}\vspace{-2mm}

    \begin{figure}[t]
    \centering
        \includegraphics[width=\linewidth]{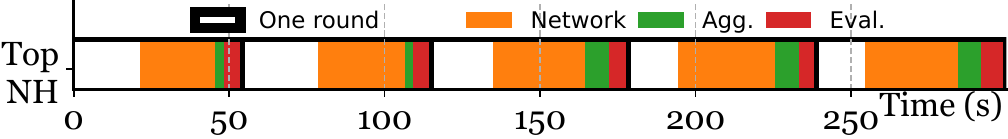}
        \includegraphics[width=\linewidth]{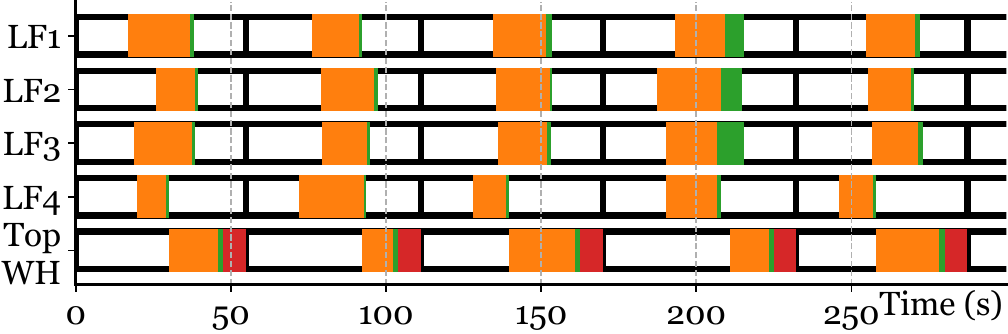}
        \vspace{-8mm}
    \caption{Impact of data plane performance on hierarchical aggregation. (upper fig.:) No hierarchy(NH); (lower fig.:) With hierarchy(WH). Top: top aggregator; LF: leaf aggregator. ``Network'' denotes the data transfer tasks of model updates; ``Agg.'' denotes the aggregation tasks; ``Eval.'' denotes the evaluation tasks.} 
    \vspace{-6mm}
    \label{fig:assess-hier-agg-dataplane}
    \end{figure}

\noindent\textbf{Assessing data plane with hierarchical aggregation:}
We now assess the importance of a high-performance data plane to truly deliver on the promise of hierarchical aggregation.  
We consider a baseline (denoted \textbt{NH}) with a single aggregator without hierarchy. 
We evaluate the hierarchical aggregation service that has one top aggregator and four leaf aggregators (denoted \textbt{WH}). All aggregators are placed on the same node.
We consider eight trainers to train a ResNet-152 model using FEMNIST dataset.
Note that we always deploy trainers on separate nodes, to both be realistic (trainers are remote) and to avoid contention for resources on the node.

Fig.~\ref{fig:assess-hier-agg-dataplane} shows the execution times for the representative FL stages under different settings.
Note that we only show the receiving part of the networking task (``Network'' in Fig.~\ref{fig:assess-hier-agg-dataplane}) to simplify the figure.
Compared to the baseline (\textbt{NH}), \textbt{WH} does not exhibit a significant improvement overall, though it uses hierarchical aggregation.
The average completion time per round with \textbt{WH} is 57 seconds, while for \textbt{NH} is 59.8 seconds.
This is mainly because of the contention for network processing  between leaf aggregators when they send/receive intermediate model updates to/from the top aggregator.
This highlights the critical need for a high-performance and streamlined data plane for  hierarchical aggregation.
\name incorporates shared memory processing  when the serverless aggregator functions are co-located on the same node. This enables fast and efficient communication, mitigating the impact of networking on hierarchical aggregation (demonstrated in Fig.~\ref{fig:hier-agg-perf}). Working jointly with our locality-aware placement scheme (\S\ref{sec:placement}), \name can minimize the need for inter-node model update transfers. Consequently, \name maximizes the advantages of our efficient intra-node shared memory data plane that substantially reduces communication overheads.

\noindent\textbf{Shared memory object store:}
The \name agent is responsible for the allocation/recycling/destruction of the shared memory buffer in the object store.
In addition, \name only allows immutable (read-only) objects to guarantee the safe sharing of model updates, eliminating the need for locks.
The agent periodically checkpoints the model parameters to an external persistent storage service (more details in Appendix-\ref{appendix:model-checkpoints}).


\vspace{-2mm}\subsection{In-place Message Queuing}\label{sec:msg-queue}\vspace{-2mm}

\begin{figure}[t]
    \centering
        \includegraphics[width=.8\columnwidth]{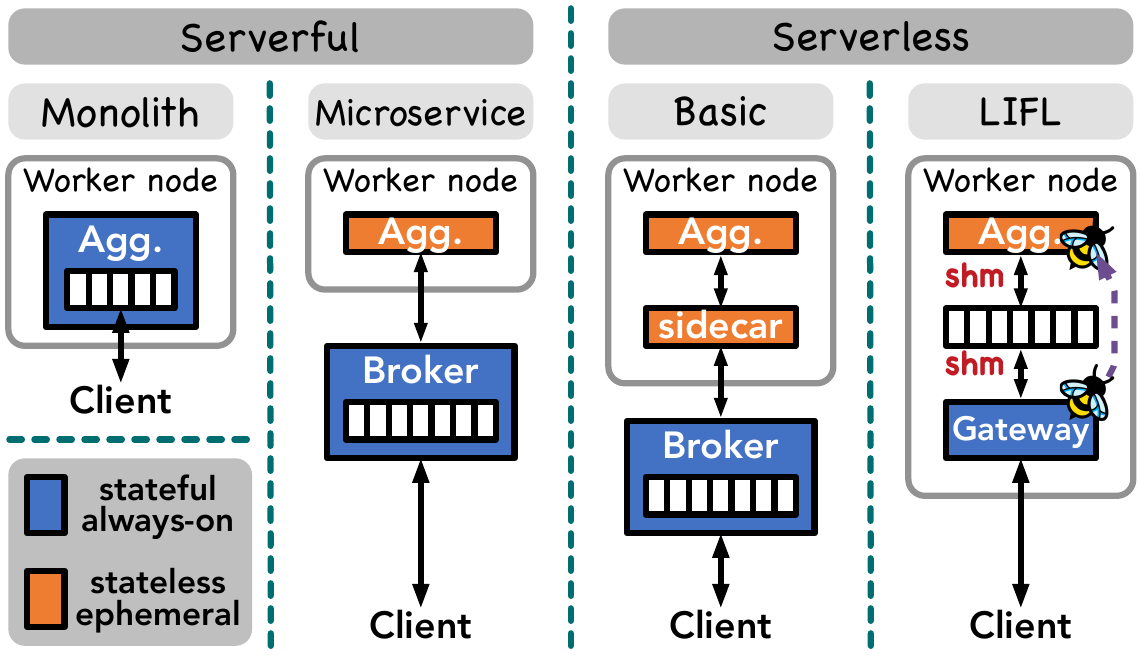}\vspace{-4mm}
    \caption{Message queuing solutions.}
    \vspace{-5mm}
    \label{fig:mq-alternatives}
    \end{figure}

\noindent\textbf{Representative message queuing solutions:}
Fig.~\ref{fig:mq-alternatives} enumerates message queuing solutions for various serverful and serverless alternatives. 
In the {\it monolithic} serverful setup (used in~\cite{papaya-mlsys2022}), the model update is directly buffered into an in-memory queue residing in the aggregator, deployed as a persistent and stateful application. 
Another serverful setup, used in~\cite{google-fl-mlsys19}, deploys aggregators as ephemeral, stateless {\it microservices}, requiring an additional persistent, stateful message broker to buffer model updates from clients before being consumed by the stateless aggregator.
Switching to the {\it basic} serverless setup (used in~\cite{adafed}), model updates are also buffered at a message broker, as the aggregator is now deployed as an ephemeral, stateless serverless function. Before being consumed by the aggregator, the model update has to pass through the sidecar.  Finally, in \name, the gateway buffers the model update directly into the shared memory, which can then be seamlessly accessed by the aggregator.
The distinct data pipelines between the client, message queue, and aggregator impose varying degrees of overheads.
Our evaluation (details in Appendix-\ref{sec:msg-queue-expt}) shows that \name's in-place message queuing achieves the best efficiency and performance (equivalent to a monolithic, serverful design) among all alternatives in Fig.~\ref{fig:mq-alternatives}.

\noindent\textbf{Message queuing pipeline in \name:}
The gateway at each worker node is addressable/accessible by FL clients. It receives model updates from clients or from the gateway on another worker node, and performs necessary network processing (\eg protocol processing, serialization, deserialization, data type conversion, {\it etc}) before writing the model updates into shared memory. This avoids duplicate processing when local aggregators access model updates in shared memory.
A step-by-step explanation of the processing flow of the message queuing in \name is given in Appendix-\ref{appendix:msg-queueing}.

We apply vertical scaling of the gateway by dynamically adjusting the number of assigned CPU cores based on the load level. This avoids the gateway becoming the dataplane bottleneck and impacting the aggregation speed.


\vspace{-2mm}\subsection{eBPF-based Sidecar}\label{sec:ebpf-sidecar}\vspace{-2mm}
\name's eBPF-based sidecar is built with a set of eBPF programs attached to each aggregator's socket interface, using its in-kernel \skmsg hook~\cite{Red2021Understanding}.
The execution of the eBPF-based sidecar is triggered by the invocation of the {\it send()} system call, which is captured by the \skmsg hook as an eBPF event. This ensures that the eBPF-based sidecar is strictly event-driven and consumes {\it no} CPU resources when idle. We use the eBPF-based sidecar to collect necessary metrics (\eg execution time of the aggregation task) to facilitate the orchestration in \name (\S\ref{sec:lifl-orches}). 


\noindent\textbf{Metrics collection:}
Upon invocation, the eBPF-based sidecar collects and stores metrics to an eBPF map ({\it metrics map})
on the local worker node.
The eBPF map is an in-kernel, configurable key-value table that can be accessed by the eBPF program during  execution~\cite{ebpf-map}.
The \name agent, on the other hand, periodically retrieves the latest metrics from the {\it metrics map} and feeds the metrics back to the metrics server (Fig.~\ref{fig:system-overview}) in the serverless control plane.

\vspace{-2mm}\subsection{Direct Routing with Hierarchical Aggregation}\label{sec:routing}\vspace{-2mm}
    
Direct networking between functions is not allowed in existing serverless environments because serverless functions are considered to be stateless and ephemeral. This implies that there are no long-lived, direct connections between a pair of function instances. As a result, they use an intermediate networking component (\eg message broker) to act as a stateful, persistent component to manage state, \ie routes between functions.
However, the main drawback is that it adds unnecessary overhead by involving the additional networking component(s) in the datapath, making indirect networking between functions heavyweight.

\name improves serverless networking within hierarchical aggregation by allowing direct routing between aggregators, both within a node and between nodes. The key is to offload the stateful processing to eBPF, using the \textit{sockmap}~\cite{Red2021Understanding} to support flexible intra-node routing exploiting shared memory, and inter-node routing with the help of the per-node gateway, as depicted in Fig.~\ref{fig:direct-routing}.
The \textit{sockmap} is a special eBPF map (\texttt{BPF\_MAP\_TYPE\_SOCKMAP}~\cite{Red2021Understanding}) that maintains references to the registered socket interfaces.
We take the approach from~\cite{spright-sigcomm22} to implement intra-node direct routing in \name. For details of intra-/inter-node routing in \name, refer to Appendix-\ref{sec:routing-details}.

\vspace{-2mm}\section{\name's Control Plane Design}\label{sec:lifl-orches}\vspace{-2mm}

\hide{\subsection{Design Implications}
\knote{I think we can remove Section 5.1 entirely. If there is anything to use from this, we can put in the specific subsections we have. All we need is the figure 8 for describing the remaining subsections,  Hierarchy Planning, Reuse, etc.}

Based our discussion in \S\ref{sec:shifting}, we summarize the considerations guiding the design of \name's orchestration heuristic:}

\hide{\textib{I1:} {\it Maximizing parallelism in each level of hierarchical aggregation.}
Adjusting the number of aggregators in each level of the hierarchy corresponds to the number of input model updates to perform aggregation tasks in parallel. Each aggregation task is represented as aggregation of a single model update. This can minimize the completion time of each level and thus the minimize the ACT of the hierarchical aggregation.}

\hide{\textib{I2:} {\it Opportunistically reusing aggregator runtime from the lower level.}
Such reuse may not be feasible for general serverless analytics use cases~\cite{nimble-nsdi21} that heterogenized function runtimes cannot be simply reused. However, it is applicable in FL because aggregators typically use homogenized runtimes with the same code and libs.
Runtime reuse
avoids the creation of additional runtime instances dedicated to higher-level aggregators, mitigating potential startup delay that may increase ACT, and improves resource efficiency by cutting costs as fewer runtime instances are used.
It is also important to timely terminate aggregator runtimes that are not being reused to avoid resource wastage.}

\hide{\textib{I3:} {\it Greedily placing aggregators from lower levels to the same node.}
This is based on the fact that the hierarchical aggregation 
typically follows a tree-based structure, in which the amount of data reduces from the lower levels to the higher levels. Partitioning aggregators in lower levels into the same placement group can maximize the benefit of high-performance shared memory processing in \name.}
\subsection{Locality-aware Placement and Load Balancing}\label{sec:placement}\vspace{-2mm}
The placement of aggregators can lead to different routing behaviors:
When aggregators with cross-level data dependencies are placed on the same node, the shared memory processing and eBPF-based sidecar can facilitate intra-node routing.
\hide{shared memory processing can be taken advantage of with the eBPF-based sidecar facilitating intra-node routing.}
When these aggregators are placed across different nodes, the gateway has to perform inter-node routing.
To minimize the transfer of model updates in \name, we take a data-centric strategy like~\cite{pheromone-nsdi} that is aware of the locality of model updates and places the aggregator close to the model updates.
As such, the in-place message queuing (\S\ref{sec:msg-queue}), which is, in fact, the result of load balancing (clients to worker node mapping), directly affects the effectiveness of the locality-aware placement of the aggregators.

Our objective of load balancing involves two crucial criteria:
\textbf{(1)} Minimizing inter-node communication
while maximizing the utilization of shared memory within each node.
\textbf{(2)} Ensuring the 
residual service capacity of the worker node meets the demand; the residual service capacity ($RC_{i,t}$) of worker node $i$ at time $t$ is determined by $RC_{i,t} = MC_{i} - (k_{i,t} \times E_{i,t})$. Here, $MC_{i}$ represents the maximum service capacity\footnote{We compute $MC_{i}$ offline; for details, refer to Appendix-\ref{sec:mc}.}, denoting the maximum number of model updates that can be  aggregated simultaneously on worker node $i$. The value of $k_{i,t}$ is the arrival rate of model updates directed to worker node $i$ at time $t$, and $E_{i,t}$ is the average execution time required to aggregate a model update
on node $i$. We can also get a coarse-grained estimate on the queue length ($Q_{i,t} = k_{i,t} \times E_{i,t}$) of node $i$ at time $t$. 


We approach the load balancing task as a bin-packing problem, aiming to 
allocate model updates from clients
to a minimal number of worker nodes, while ensuring that
the residual service capacity of each worker node is not exceeded. This naturally reduces the inter-node communication as much as possible, since the communication between a particular pair of worker nodes only happens once. We use \textit{BestFit} for the bin-packing, as it concentrates load onto the fewest nodes possible, to reduce inter-node traffic and maximize shared memory use. In contrast, \textit{WorstFit} spreads the load across more nodes, similar to the ``Least Connection'' policy in Knative (\S\ref{sec:micro}). Furthermore, \textit{FirstFit} focuses on reducing search complexity without being locality-aware.

    \begin{figure}[t]
    \centering     
        \includegraphics[width=\columnwidth]{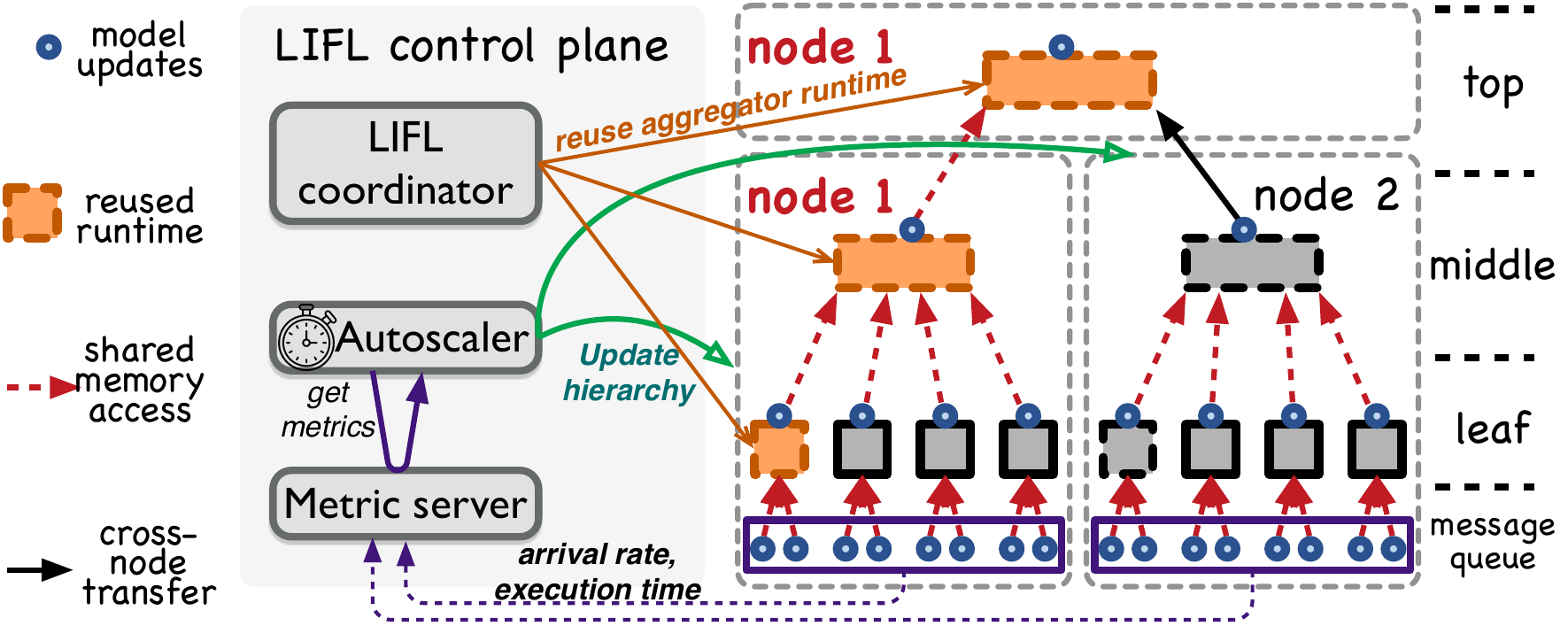}
        \vspace{-7mm}
    \caption{Control plane orchestration in \name: The autoscaler periodically re-plans the hierarchy based on the arrival rate of each worker node. The \name coordinator applies reusing of aggregators.}
    \vspace{-7mm}
    \label{fig:pareto-orches}
    \end{figure}

\vspace{-2mm}\subsection{Planning the Hierarchy for Aggregation}\label{sec:hier-plan}\vspace{-2mm}





The goal of hierarchy-aware autoscaling is to maximize the parallelism of aggregation at each level, given the number of model updates to be aggregated.
This can minimize the completion time of each level and thus minimize the aggregation completion time (ACT) for hierarchical aggregation.
We plan a hierarchical aggregation structure within each node, tailored to the number of pending model updates ($Q_{i,t}$) in the message queue.
Every node produces an intermediate model update that is dispatched to the node chosen to have the top aggregator that updates the global model.
This approach significantly reduces the need for cross-node transfers for intermediate model updates.

\name periodically adjusts (\ie scales) the hierarchy on node $i$, guided by our estimates of $Q_{i,t}$.
To prevent excess resource allocation due to short-term spikes in $Q_{i,t}$, we employ the Exponentially Weighted Moving Average (EWMA) to smooth $Q_{i,t}$: $Q_{i,t} = \alpha\times Q_{i,t-1} + (1- \alpha)\times Q_{i, t}$, where $\alpha$ is the EWMA coefficient. We set $\alpha = 0.7$ based on it yielding the best results in our experiments.
Our current implementation supports a two-level \textit{k}-ary tree hierarchy on each node, comprising a ``central'' middle aggregator responsible for aggregating model updates from $Q_{i,t}/I$ leaf aggregators, where $I$ is the number of model updates of clients per leaf aggregator.
Given that the steps within a \name aggregator (Fig.~\ref{fig:step-model}) are executed sequentially, we want to maximize the parallelism by having a limited $I$ to be small (\eg at 2), ensuring that a leaf aggregator experiences minimal waiting time after receiving the initial update from the first client.

\name re-plans the hierarchy on each worker node
periodically. 
This involves estimating 
$Q_{i,t}$ across the worker nodes 
and creates/terminates aggregators accordingly. The \name control plane updates the routes between aggregators based on the renewed hierarchy (details in Appendix-\ref{sec:routing-details}).

\hide{Based on the processing model in \S\ref{sec:proc-model}, with each aggregator handling model aggregation of {\color{red}k} model updates at a time, the hierarchical aggregation in \name can be viewed as a k-ary tree~\cite{}, as depicted in Fig.~\ref{fig:pareto-orches}. 

Depending on the parity of $M$, the binary tree is {\it full} when $M$ is even, and is {\it complete} when $M$ is odd\sxnote{reference?}.
We generate the binary tree based on the set of $M$ model updates as leaf nodes, using the hierarchy planning algorithm in Algorithm~\ref{algo:generate-binary-tree}. This generates a balanced binary tree 
with $\lceil \log_2 M \rceil$ levels in total and $2^i$ nodes at level $i$.}

\vspace{-3mm}\subsection{Opportunistic Reuse of Aggregator Instances}\label{sec:reuse}\vspace{-2mm}

The scaling policy in \name incorporates an opportunistic ``reuse'' scheme to maximize the utilization of warm aggregator instances since aggregators in \name 
use homogenized runtimes (Fig.~\ref{fig:step-model}) with the same code and libs. 
This sidesteps the cascading effect~\cite{graf-conext} when starting up a hierarchy of aggregators (in fact function chains).

Given a hierarchy of aggregators selected on the node, \name picks a leaf aggregator that has already completed its aggregation task and is idle. \name converts its role to a middle aggregator on that node. No further change is required as \name's aggregator runtime is stateless.
\name selects the first middle aggregator that completes its local aggregation task and converts it to be the top aggregator responsible for updating the global model.
This minimizes the need to start up new instances for higher-level aggregators, and avoids additional startup delays.





\vspace{-3mm}\subsection{Eager aggregation in \name}\label{sec:eager}\vspace{-2mm}


\name employs eager aggregation (Fig.~\ref{fig:sync-async-fl}) leveraging its more flexible and dynamic timing of the aggregation process. Eager aggregation performs timely aggregation as model updates arrive, even if it triggers the cold start of an aggregator (when no idle-but-warm aggregator is available). This takes advantage of the {\it overlap} between the start-up delay and transfers of model updates, allowing eager aggregation to mask cold starts up until the last model update. It also mitigates congestion that can occur when trying to aggregate all model updates simultaneously. In contrast, lazy aggregation aggregates all model updates in a batch when the aggregation goal is reached. But, the arrival of local model updates from trainers can be spread over a relatively long duration.
Our evaluation shows eager aggregation achieves a 20\% reduction on ACT (Fig.~\ref{fig:act-optimality}).
We implement eager aggregation in \name following the step-based processing model described in Appendix-\ref{sec:step-model}.
\name updates the version of the global model whenever the aggregation goal is achieved.



\hide{\subsection{Additional Considerations}\vspace{-2mm}
\noindent\textbf{Client selection:}
\name can seamlessly work with various advanced client selection schemes (\eg Oort~\cite{oort}, REFL~\cite{refl}, Pisces~\cite{pisces-socc22}), etc., by properly configuring the coordinator. The coordinator partitions selected clients across selectors on different worker nodes. The client establishes a connection with the assigned selector to upload local updates.}

\hide{\noindent\textbf{Startup latency of aggregator:}
Running the model aggregation as serverless functions can contribute to delay waiting for the additional function(s) to startup. 
Prior work examines the mitigation of function startup delay through ``just-in-time'' scheduling in various serverless use cases, \eg serverless analytics~\cite{nimble-nsdi21}, and serverless FL aggregation~\cite{jit-agg}.

Serverless analytics workloads depend on significant deterministic behavior based on input job information and leverage static call graphs between tasks that are described in a DAG~\cite{nimble-nsdi21}. Using ``just-in-time'' scheduling, they can mitigate the impact of startup latency.
But, unlike serverless analytics, FL aggregation can be much more dynamic, \ie when using hierarchical aggregation, the hierarchy has to be dynamically scaled in response to the instantaneous load level; the arrival time of model updates from trainers is also likely to be more random, reflecting the varying execution times ad trainers.
Although \citet{jit-agg} propose a prediction-based approach to enable ``just-in-time'' scheduling of serverless aggregation,
it is difficult to precisely predict the arrival timing of the update from {\it every single client} under a large-scale setup, serving large numbers of trainer devices. 

Our evaluation shows that \name's hierarchy planning and reuse scheme can avoid the startup delay by keeping aggregators warm  (\S\ref{sec:micro}), without impacting the convergence rate (\S\ref{sec:system-eval}).}

\hide{\noindent\textbf{Load balancing (client to leaf aggregator mapping):}
Due to the stateless nature of \name's aggregator, there is no fixed binding between clients and leaf aggregators. 
With \name's in-place message queuing, FL clients are randomly assigned to a leaf aggregator. This further enables secure aggregation, which enhances privacy of client data, to help prevent the aggregation process from learning about the individual client's data.}






\vspace{-3mm}\section{Evaluation \& Analysis}\label{sec:evaluation}\vspace{-2mm}

We quantify the performance gain and resource savings by using \name, starting with analyzing a set of microbenchmarks to understand the different design considerations of \name, including shared memory processing, the effectiveness and overheads of \name's orchestration scheme.
We then demonstrate the benefits of \name from a system-level perspective using real FL workloads.
\hide{We seek to answer the following questions through the evaluation: \\
(1) Does \name use resources more efficiently compared to existing serverful and serverless solutions? \\
(2) Is there a tradeoff in \name between resource efficiency and FL performance, \eg convergence rate? \\
(3) Does \name's optimizations impact  model accuracy?}

\noindent\textbf{Baseline Systems:}
We implement several baseline FL systems for \name to compare against.
\textbf{(1) ``Serverful system'' (SF):}
The ``serverful system'' is implemented following the design described in~\cite{google-fl-mlsys19} and~\cite{papaya-mlsys2022}. Both of them adopt the architecture depicted in Fig.~\ref{fig:fl-system} (a).
\textbf{(2) ``Serverless system'' (SL):} 
The baseline ``serverless system'' is implemented following the design described in FedKeeper~\cite{fedkeeper} and AdaFed~\cite{adafed} that uses the architecture depicted in Fig.~\ref{fig:fl-system} (b). We choose Knative~\cite{knative} as the serverless framework to build these alternatives. 
We utilize the open-source Flame platform~\cite{flame-cisco} to provide necessary FL components, \eg coordinator, selector, aggregator, and client.

\hide{We consider two variants for this ``serverless'' baseline. The first variant (denoted \textbt{SL}) adopts the design of FedKeeper~\cite{fedkeeper} that does not consider hierarchical aggregation support, using at most one aggregator to handle all model updates. The second variant (denoted \textbt{SL-H}) adopts the design of AdaFed~\cite{adafed}, using hierarchical aggregation to support FL at scale.} 

\noindent\textbf{Implementation of \name:}
We implement \name based on SPRIGHT~\cite{spright-sigcomm22}, a lightweight, high-performance serverless framework.
\name includes object store support, model checkpoints, and routing support for hierarchical aggregation. \name uses Python's multiprocessing package to implement the shared memory pool instead of the DPDK-based shared memory pool used in the original implementation of SPRIGHT.
The current implementation of \name only supports synchronous FL. Supporting asynchronous FL is part of our future work.

\textib{Testbed setup:}
We leverage the NSF Cloudlab~\cite{cloudlab}. The nodes we used have a 64-core Intel Cascade Lake CPU@2.8 GHz, 192GB memory, and a 10Gb NIC. We use Ubuntu 20.04 with kernel version 5.16.

\begin{figure}[t]
    \centering
        \subfigure[latency of a single model update transfer (intra-node)]{
            \includegraphics[width=.47\columnwidth]{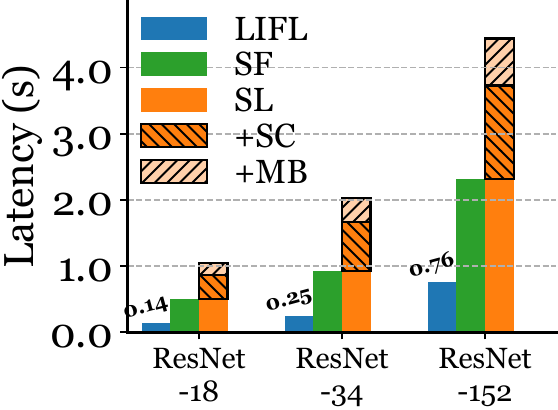}
            \label{fig:hier-latency}
        } \hfill 
        \subfigure[CPU usage of a single model update transfer (intra-node)]{
            \includegraphics[width=.47\columnwidth]{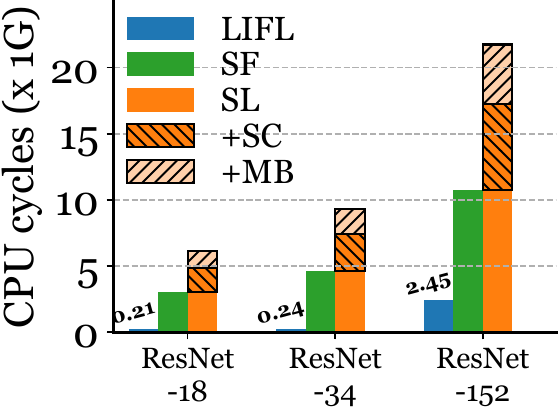}
            \label{fig:hier-cpu}
        } 

    \includegraphics[width=0.6\columnwidth]{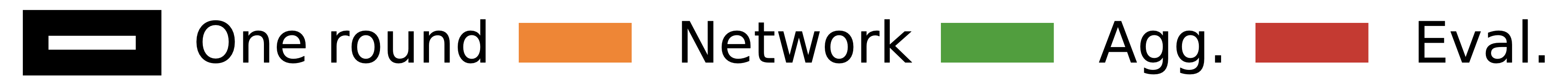}\vspace{-2mm}

        \subfigure[\name's Aggregation Timing (with ResNet-152)]{
            \includegraphics[width=\columnwidth]{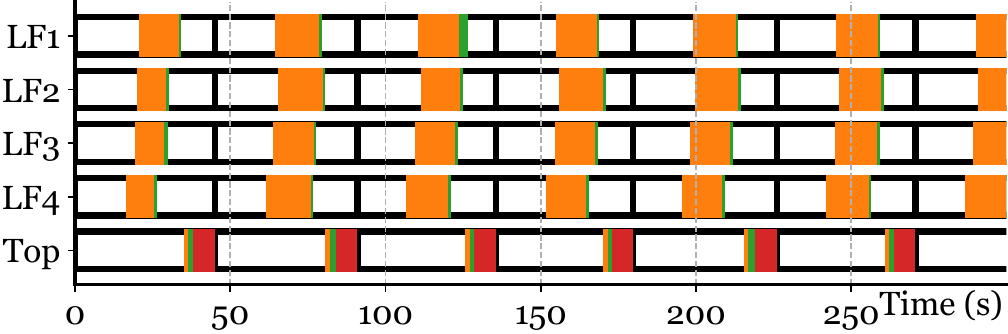}
            \label{fig:hier-timing}
        } 
        \vspace{-6mm}
  \caption{Data plane improvement for hierarchical aggregation: Serverful (\textbt{SF}), Serverless (\textbt{SL}), and \name. \textbt{SL}'s latency includes contributions of \textbt{+SC} (sidecar) and \textbt{+MB} (message broker).}
  \vspace{-8mm}
  \label{fig:hier-agg-perf}
\end{figure}

\vspace{-3mm}\subsection{Microbenchmark Analysis}\label{sec:micro}\vspace{-2mm}

\noindent\textbf{Data plane improvement for hierarchical aggregation:}
To understand the improvements in data plane performance of hierarchical aggregation with \name's shared memory processing, 
we use the same aggregation hierarchy as in \S\ref{sec:shm-support}, comprising one top aggregator and four leaf aggregators.
All aggregators are placed on the same node.

We consider the following serverful and serverless alternatives: (1) The serverful setup (\textbt{SF}) establishes direct networking
channels (based on gRPC) between leaf aggregators and the top aggregator;
(2) the serverless setup (\textbt{SL}) uses indirect networking to connect leaf aggregators and the top aggregator, through a message broker on the same node. Each aggregator has a container-based sidecar to mediate inbound and outbound traffic;
(3) the \name setup uses shared memory for communication between aggregators.
We consider three ML models with distinct sizes: ResNet-18 ($\sim$44MB), ResNet-34 ($\sim$83MB), and ResNet-152 ($\sim$232MB).

Fig.~\ref{fig:hier-latency} shows the latency breakdown of a single model update transfer between the leaf aggregator and top aggregator for different model sizes. We specially mark the share of sidecar (\textbt{+SC}) and message broker (\textbt{+MB}) for the serverless setup.
\textbt{SL} consistently results in 2\X and 6\X higher latency than \textbt{SF} and \name, respectively. 
The significant CPU usage of \textbt{SL} (Fig.~\ref{fig:hier-cpu})
clearly shows the poor efficiency and performance of the indirect networking used in the serverless setup, caused by its use of the message broker and heavyweight sidecar.
We see that \name is considerably better than \textbt{SF} and \textbt{SL} in terms of both CPU usage and latency.

Fig.~\ref{fig:hier-timing} shows the timing of various FL processing tasks during hierarchical aggregation when using \name's data plane.
It is clear that \name's shared memory processing helps reduce the overhead and improve the performance of the data plane with hierarchical aggregation.
\name completes each round in just 44.9 seconds compared to 57 seconds on average even for the serverful setup in Fig.~\ref{fig:assess-hier-agg-dataplane}.
Further, through careful placement, aggregators in \name can fully exploit the high-speed intra-node data plane over shared memory, as discussed next.


\hide{As shown in Fig.~\ref{fig:p2p-latency-intra} and~\ref{fig:p2p-cpu-intra}, for different model sizes,
\textbt{SL} consistently achieves {\color{red} XX\%} higher latency than \textbt{SF} and \name. At the same time, it uses {\color{red} XX\% to XX\%} more CPU. 
This clearly shows the poor efficiency and performance of the indirect networking used in serverless setup, \new{caused by its use of the message broker and heavyweight sidecar.} 
\hide{We further quantify the CPU and latency specifically incurred by the message broker and sidecar in \textbt{SL}. The result shows that the message broker accounts for {\color{red} XX\%} of the total overhead (both CPU and latency), and the sidecar accounts for {\color{red} XX\%} of the total overhead, which is also quite significant.
Although \name involves the gateway to support inter-node communication between aggregators, the communication between the local aggregator and the gateway is through shared memory (Fig.~\ref{fig:direct-routing}), incurring negligible overhead.
The major dataplane overhead comes from the communication between gateways on the different worker nodes, which is equal to that of the direct networking between aggregators using the \textbt{SF} framework.
The result is that the inter-node dataplane performance of \name is quite close to \textbt{SF}.}
We see that \name is significantly better than \textbt{SF} and \textbt{SL} in terms of both CPU usage and latency. This clearly shows the benefit of utilizing shared memory 
to help reduce the overhead and improve the performance of the dataplane with hierarchical aggregation. 
Further, through careful placement, aggregators in \name can fully exploit the high-speed intra-node dataplane over shared memory, as discussed next.}

\hide{In this evaluation of the intra-node dataplane, all 3 aggregators are placed on the same node. We also place the message broker on the same node as aggregators with the serverless setup. \name setup uses the {\it intra-node} shared memory for communication (Fig.~\ref{fig:direct-routing}) this time. We see that \name is significantly better than \textbt{SF} and \textbt{SL} in terms of both CPU usage and latency. This clearly shows the benefit of utilizing shared memory (leveraged by \name's locality-aware placement) to help reduce the overhead and improve the performance of the dataplane with hierarchical aggregation. The \textbt{SL} still suffers from the heavyweight dataplane caused by the use of indirect networking.}

\begin{figure}[t]
  \centering
        \subfigure[Agg. Completion Time]{
            \includegraphics[width=.47\columnwidth]{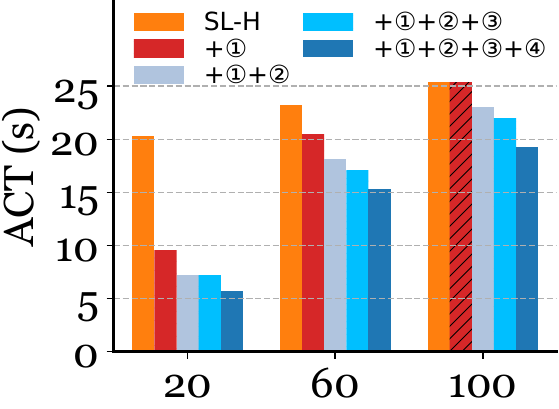}
            \label{fig:act-optimality}
        }
        \subfigure[Cumulative CPU Time]{
            \includegraphics[width=.47\columnwidth]{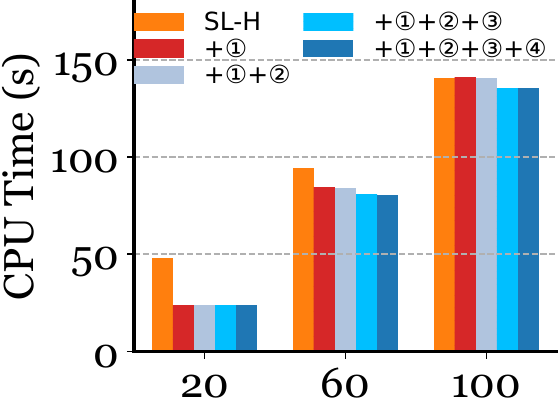}
            \label{fig:cost-optimality}
        }
        \subfigure[\# of aggregators created]{
            \includegraphics[width=.47\columnwidth]{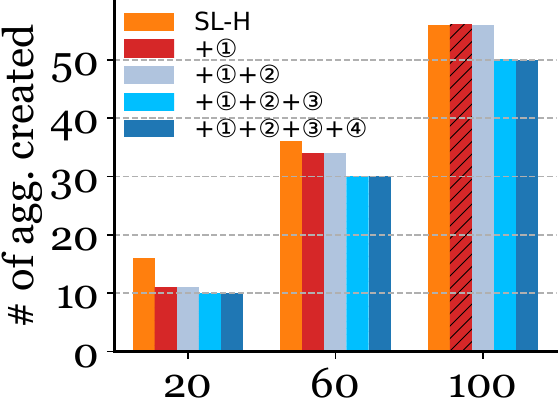}\vspace{-4mm}
            \label{fig:agg-created}
        }
        \subfigure[\# of nodes used]{
            \includegraphics[width=.47\columnwidth]{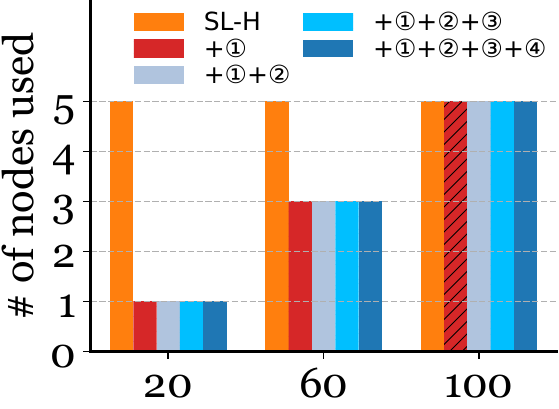}
            \label{fig:node-used}
        }
        \vspace{-6mm}
  \caption{Improvement with \name's orchestration, with \circlednumber{i} being additions to baseline \name; {\it x-axis is the number of model updates arriving at the aggregation service concurrently.}}
  \label{fig:cost-act-optimality}
  \vspace{-6mm}
\end{figure}

\noindent\textbf{Improved orchestration in \name:}
We now quantify the benefits of \name's orchestration in improving hierarchical aggregation.
We demonstrate the effectiveness of \name
by applying:  \circlednumber{1} locality-aware placement (\S\ref{sec:placement}), \circlednumber{2} hierarchy-planning (\S\ref{sec:hier-plan}), \circlednumber{3} aggregator reuse (\S\ref{sec:reuse}), and \circlednumber{4} eager aggregation (\S\ref{sec:eager}) step-by-step.
We use five nodes for this experiment. The maximum service capacity ($MC_i$) of each node in our testbed is 20.\footnote{Our testbed nodes are homogeneous, hence all $MC_i$ are the same. With heterogeneous nodes, $MC_i$ may vary.}
We focus on two aspects: resource consumption and Aggregation Completion Time (ACT) to aggregate a given number of model updates. In this experiment, we assume the estimated $Q_{i,t}$ is equal to the actual queue length on each active node. We focus on the importance of having warm aggregators based on the pre-planned hierarchy, to avoid the cold start penalty.

We compare \name against a baseline serverless control plane using hierarchical aggregation (\textbt{SL-H} in Fig.~\ref{fig:cost-act-optimality}). \textbt{SL-H} employs \name's shared memory data plane (so both have the same data plane) with Knative's ``Least Connection'' load balancing strategy~\cite{mu-socc21} that assigns newly arrived model updates to the node with the smallest queue length.
The aggregators in \textbt{SL-H} use lazy aggregation by default. The ML model used is ResNet-152. Note that the latency to transmit a single model update of ResNet-152 across nodes (on the current testbed) is $\sim$4.2 seconds.


By using locality-aware placement,  \name also achieves 2.1\X and 1.13\X ACT reduction than \textbt{SL-H} (for 20 and 60 model updates in Fig.~\ref{fig:act-optimality}). This improvement is attributed to \name's bin-packing strategy, which effectively consolidates aggregators onto the same node to fully exploit shared memory processing.
Applying hierarchy-planning and reusing warm aggregator instances (+\circlednumber{1}+\circlednumber{2}+\circlednumber{3}) further reduce $\sim$1.22\X ACT of \name, as keeping aggregators warm mitigates the cold start delay that exists in both \textbt{SL-H} and (+\circlednumber{1}).
Further, after enabling eager aggregation (+\circlednumber{1}+\circlednumber{2}+\circlednumber{3}+\circlednumber{4}), \name allows higher-level aggregators to consume and aggregate the model updates in a timely manner, effectively avoiding the intermediate model updates (produced by the lower-level aggregators) being queued up at the higher-level aggregators. This saves $\sim$1.2\X in ACT compared to (+\circlednumber{1}+\circlednumber{2}+\circlednumber{3}) that uses lazy aggregation.

While being effective in reducing ACT, \name also helps to reduce costs. Just using  locality-aware placement (+\circlednumber{1} in Fig.~\ref{fig:cost-optimality}), \name can save considerable CPU overhead by reducing inter-node data transfers (with 20 and 60 model updates). Enabling aggregator reuse saves additional CPU cycles, as it avoids having the CPU initialize new aggregators. For 100 model updates though, the service capacity of all five nodes would be maxed out, reaching the limit of the benefit of \name's orchestration. However, the data plane improvement of \name can still make it outperform the basic serverful and serverless setups, as demonstrated in Fig.~\ref{fig:hier-agg-perf}.

As shown in Fig.~\ref{fig:agg-created}, \name reduces the number of aggregators created, by packing more aggregators into fewer nodes.
After we apply locality-aware placement to \name (+\circlednumber{1}), \name can also reduce the number of nodes used considerably (see Fig.~\ref{fig:node-used}): 
Given 20, 60, and 100 model updates, \name's locality-aware placement efficiently packs them into 1, 3, and 5 nodes, respectively.
This avoids repeatedly creating a middle aggregator on each of the 5 nodes (except when the service capacity of all 5 nodes is fully consumed). 
On the other hand, \textbt{SL-H} uses all 5 nodes throughout, 
uniformly distributing model updates across all 5 available nodes. This will lead to additional cross-node data transfers, regardless of available model updates.
Note that the service capacity of all 5 nodes is fully consumed for 100 model updates.


\hide{When hierarchy-planning is applied with \name (\circlednumber{1} in Fig.~\ref{fig:cost-act-optimality}), \name's ACT is {\color{red} XX} less than \textbt{SL-H}.
However, because hierarchy-planning seeks to maximize parallelism, \name's resource allocation is {\color{red} XX} higher than the baseline. After reusing the aggregator instance (\circlednumber{1}$+$\circlednumber{2} in Fig.~\ref{fig:cost-act-optimality}), \name's resource allocation is reduced and is  {\color{red} XX} less than the baseline. At the same time, \name's ACT is unaffected.
The ACT of \name is further reduced by {\color{red} XX} when applying locality-aware placement (\circlednumber{1}$+$\circlednumber{2}$+$\circlednumber{3} in Fig.~\ref{fig:cost-act-optimality}), which benefits from the exploitation of the intra-node shared memory processing.
Overall, \name's ACT is {\color{red} XX} less than the baseline, and its resource allocation is {\color{red} XX} less than the baseline, clearly demonstrating the improvement in \name's orchestration design.}


\noindent\textbf{Orchestration overhead of \name:}
We evaluate the orchestration overhead of \name, given a different number of clients.
The time for completing the locality-aware placement in \name
is less than 17 milliseconds, even with 10K clients, which is the maximum number of client settings observed in Google's production FL stack~\cite{google-fl-mlsys19}.
Compared to the ACT, which takes several tens of seconds with a large amount of clients, this overhead for locality-aware placement is negligible. 
The EWMA estimator for hierarchy-planning takes 0.2 milliseconds per estimate, which is also negligible compared to the 2-minute cycle time used by \name to re-plan the hierarchy on each worker node.
The aggregator reuse and eager aggregation incur almost no overhead, as they do not require active involvement of the \name control plane.


    \begin{figure*}[htbp]
    \centering
        \includegraphics[width=.5\columnwidth]{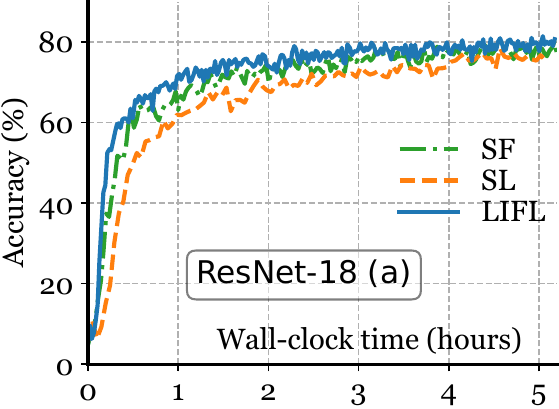}\hfill
        \includegraphics[width=.5\columnwidth]{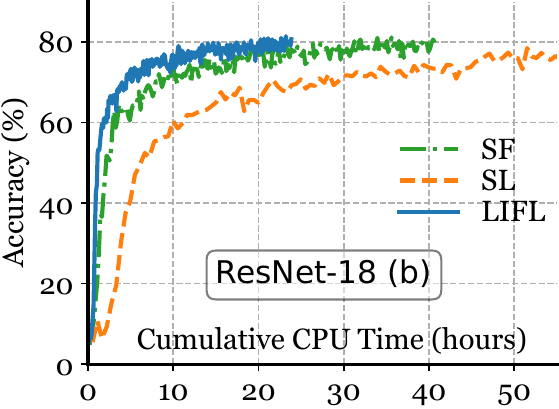}
        \includegraphics[width=.5\columnwidth]{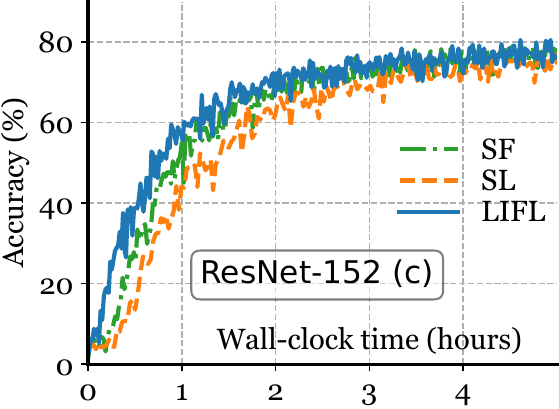}\hfill
        \includegraphics[width=.5\columnwidth]{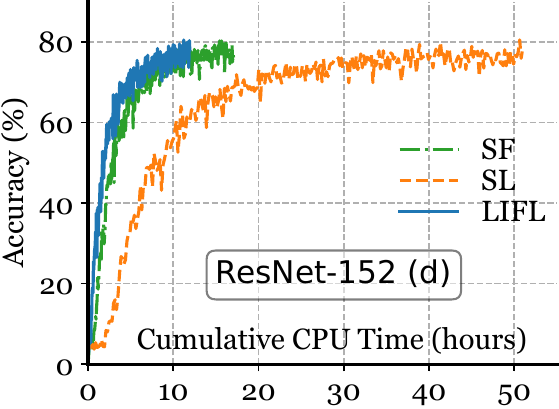}
    \vspace{-4mm}
    \caption{{\bf ResNet-18:} (a) Time-to-accuracy and (b) Cost-to-accuracy; {\bf ResNet-152:} (c) Time-to-accuracy and (d) Cost-to-accuracy.} 
    \vspace{-3mm}
    \label{fig:resnet-18}
    \end{figure*}


 \begin{figure*}[htbp]
    \centering
        \subfigure[Update arrival rate]{
            \includegraphics[width=.3\linewidth]{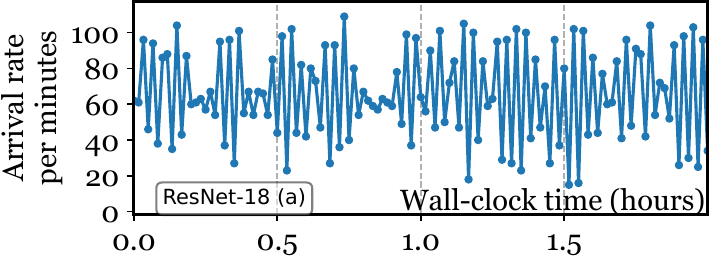}
            \label{fig:arrival-rate-resnet18}
        } \hfill
        \subfigure[\# of active aggregators]{
            \includegraphics[width=.3\linewidth]{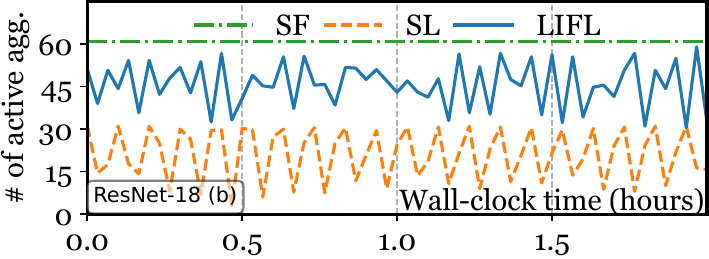}
            \label{fig:num-used-agg-resnet18}
        } \hfill
        \subfigure[Cumul. CPU time (seconds) per round]{
            \includegraphics[width=.3\linewidth]{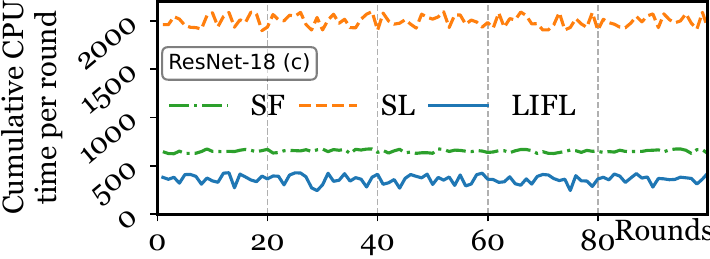}
            \label{fig:cpu-time-rounds-resnet18}
        } \vfill
        \subfigure[Update arrival rate]{
            \includegraphics[width=.3\linewidth]{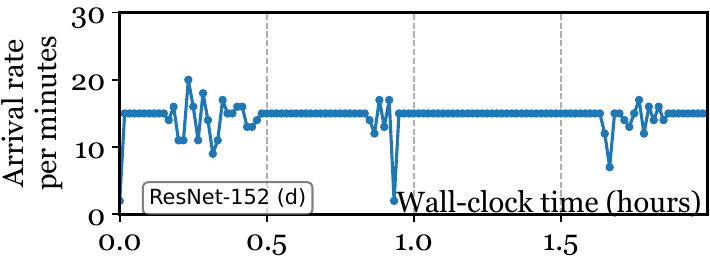}
            \label{fig:arrival-rate-resnet152}
        } \hfill
        \subfigure[\# of active aggregators]{
            \includegraphics[width=.3\linewidth]{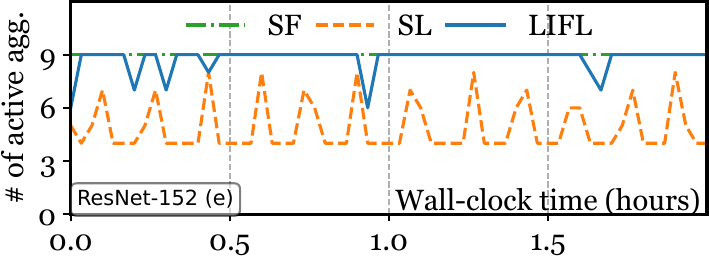}
            \label{fig:num-used-agg-resnet152}
        } \hfill
        \subfigure[Cumul. CPU time (seconds) per round]{
            \includegraphics[width=.3\linewidth]{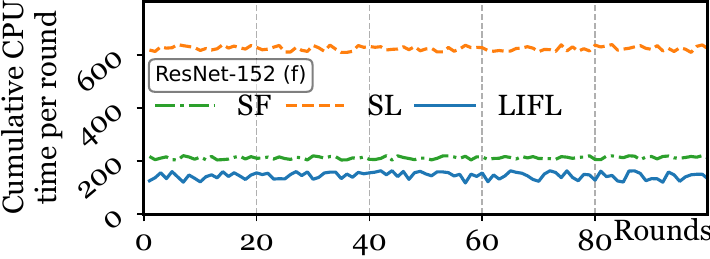}
            \label{fig:cpu-time-rounds-resnet152}
        }
        \vspace{-4mm}
    \caption{{\bf ResNet-18 (a, b, c), ResNet-152 (d, e, f):} Time series of arrival rate, number of active aggregators, and cumulative CPU time (seconds) per round.}
    \vspace{-5mm}
    \label{fig:dynamics}
    \end{figure*}


    

\vspace{-3mm}\subsection{FL Workloads Setup}\label{sec:workload}\vspace{-2mm}
\hide{We focus on two typical FL scenarios for evaluation: {\it cross-device FL} (e.g., for user devices, such as mobile phones) and {\it cross-enterprise (or cross-silo) FL} (e.g., for enabling collaboration among organizations, with more capable client devices).}
Our aim is to demonstrate the generality of \name in improving performance and reducing the cost of FL from a system-level perspective. 
We consider synchronous FL (using FedAvg~\cite{fedavg}) to justify \name's design. 
We use Stochastic Gradient Descent on the client. Clients are configured with a batch size of 32 in a local training epoch, with the learning rate set to 0.01.

\textib{Benchmark selection:}
We consider image classification, training ResNet~\cite{resnet} models with the FEMNIST dataset~\cite{medmnistv1}. 
We use non-IID datasets from FedScale~\cite{fedscale} (with its real client-data mapping) to keep the setting realistic with different data distributions across the client population.


\hide{\textib{FL algorithm:}
We consider both synchronous FL and asynchronous FL to justify their impacts on \name's design. We use FedAvg~\cite{fedavg} for aggregation in synchronous FL and FedBuff~\cite{fedbuff} for aggregation in asynchronous FL. Under both setups, we use SGD on the client. We configure the client to use a batch size of 32 in a local epoch of training. The learning rate is set to 0.01.}




\textib{Configuration of clients:}
We consider two distinct client setups: 
(\textib{ResNet-18} setup) We use the client in this setup to train a ResNet-18 model. Clients are considered to be mobile devices with limited computing capacity, available only when each has battery power and is connected to a data (\eg WiFi) network. This results in high variability in the number of mobile devices available to perform training tasks. As such, we let each client hibernate for a random interval within [0, 60] seconds to emulate the dynamic availability of typical mobile device behavior. This generates varying loads over time, as shown in Fig.~\ref{fig:arrival-rate-resnet18}, justifying the need for scaling with a serverless framework as well as \name.
(\textib{ResNet-152} setup) The client in this setup trains the relatively heavyweight ResNet-152 model. The client is considered to be a server with substantial computing capacity and is highly available. As such, we keep clients in this setup always-on. This results in a more stable arrival pattern of model updates, as shown in Fig.~\ref{fig:arrival-rate-resnet152}.

We use a total of 20 physical nodes with 5 nodes used to run aggregators. We use 4 nodes as leaf/middle aggregators and dedicate one node to be the top aggregator. To deliver the benefits of a ``serverful system'' (\textbt{SF}), we always maximize the resource allocation to the aggregators and keep them warm throughout the experiment. For the serverless setup (\textbt{SL} and \name), we create aggregators on demand.

We use the remaining 15 physical nodes to run the clients. In the \textib{ResNet-18} setup, since we consider clients to be compute-constrained mobile devices, we run eight clients on the same physical node, so each client only gets a small share of the compute capacity of the physical node. 
Therefore, in the \textib{ResNet-18} setup, we can keep 120 simultaneously active clients in each round.
In the \textib{ResNet-152} setup, we treat the client as a server node, so we dedicate a physical node for a ResNet-152 client.
In this \textib{ResNet-152} setup, we keep 15 simultaneously active clients in each round. The active clients are selected from a total of 2,800 real clients provided by FedScale~\cite{fedscale}.

\vspace{-3mm}\subsection{Putting It All Together}\label{sec:system-eval}\vspace{-2mm}


\noindent(\textib{ResNet-18}) \textbf{Time to Accuracy:}
We compare the time-to-accuracy 
of \name against \textbt{SL} and \textbt{SF}. To reach 70\% accuracy of ResNet-18 (Fig.~\ref{fig:resnet-18} (a)), \name takes only 0.9 hours (wall clock time), which is 1.6\X faster than \textbt{SF} (1.4 hours). Compared to \textbt{SL} which takes 2.4 hours, \name is 2.7\X faster.
The improvement with \name can be attributed to the shared memory data plane and the improved orchestration to effectively utilize resources, thereby reducing ACT (see \S\ref{sec:micro}).


The time spent by the \textbt{SL} aggregation service increases due to a combination of factors including sidecar overhead, function chaining, and simplistic orchestration. Frequent start-up of the aggregators in \textbt{SL} (Fig.~\ref{fig:num-used-agg-resnet18}) adds delays to the aggregation (for the first arrival update in a round).
This increased aggregation time of \textbt{SL} eventually hurts the time-to-accuracy (70\%), making it even slower than \textbt{SF}.

\hide{On the other hand, SL-H, for its use of hierarchical aggregation, greatly alleviates the congestion built at the message queue and brings more elasticity into the aggregation service to dynamically adapt to changes in load. However, due to the entrenched overhead of existing serverless frameworks, the time spent on aggregation service increases due to a combination of factors including the sidecar, function chaining, the dataplane delay. This makes the convergence rate of \textbt{SL-H} {\color{red} XX} slower than \name.}

\hide{The comparison with \textbt{SL} and \textbt{SL-H} clearly demonstrates the advantages of \name with its optimized serverless infrastructure plus more fine-grained resource orchestration. }


\noindent(\textib{ResNet-18}) \textbf{Cost savings with \name:}
\name achieves significant cost savings compared to \textbt{SF} and \textbt{SL}.
We focus on the cumulative costs (CPU time) consumed by the aggregation service to achieve a certain model accuracy. 
To reach the 70\% accuracy level of ResNet-18 (Fig.~\ref{fig:resnet-18} (b)), \name consumes 4.5 CPU hours, which is 1.8\X less than \textbt{SF} (8 CPU hours). 
Further, \textbt{SF}, with its simplistic fixed resource allocation, keeps aggregators ``always-on'', constantly occupying its CPU allocation (Fig.~\ref{fig:num-used-agg-resnet18}).
\name adapts well to the arrival rate of model updates and re-plans (scales) the hierarchy accordingly, using resources to match demand.
Also note that the \name's aggregator, when deployed as a Kubernetes pod or container, is also cheaper (smaller resource allocation) than \textbt{SF}, as \name requires less CPU to complete the same amount of aggregation tasks (Fig.~\ref{fig:cpu-time-rounds-resnet18}).


In contrast, \textbt{SL} consumes much more CPU (26 CPU hours) to achieve the 70\% accuracy level of ResNet-18 (Fig.~\ref{fig:resnet-18} (b)) compared to \name (4.5 CPU hours).
Although \textbt{SL} has relatively fewer active aggregators over time (Fig.~\ref{fig:num-used-agg-resnet18}), 
its data plane and sidecar overheads, and the CPU consumed for start-up results in \textbt{SL} having more than 5\X the CPU consumption of \name. This higher CPU time cost per round (for the same amount of aggregation work completed) requires the cloud service provider to allocate far more resources to the aggregator (\eg as a pod), making a single aggregator in \textbt{SL} much more expensive than both \textbt{SF} and \name.


\noindent(\textib{ResNet-152}) \textbf{Time to Accuracy:}
Fig.~\ref{fig:resnet-18} (c) shows the time-to-accuracy of the different alternatives for RestNet-152. To reach 70\% accuracy, \name takes 1.9 hours (wall clock time), which is 1.15\X faster than \textbt{SF} (2.2 hours). Comparatively, \textbt{SL} takes 3.2 hours. \name is 1.68\X faster than \textbt{SL}.
The heavy-weight sidecar, slow function chaining, function startup delays, and simplistic orchestration, are responsible for the larger time-to-accuracy of \textbt{SL} for ResNet-152, just as we saw with the ResNet-18 workload, as well as with the microbenchmark analysis.


\noindent(\textib{ResNet-152}) \textbf{Cost savings with \name:}
As Fig.~\ref{fig:resnet-18} (d) shows, \name again achieves significant cost savings (on cumulative CPU time) compared to \textbt{SF} and \textbt{SL}.
To reach the 70\% accuracy level of the ResNet-152 model, \name consumes 4.76 CPU hours, which is 1.43\X less than \textbt{SF} (6.81 CPU hours).
In contrast, \textbt{SL} consumes much more CPU (20.4 CPU hours) to achieve the same 70\% accuracy level compared to \name.
This again is consistent with what we observed from the ResNet-18 workload, highlighting the advantage of \name.

\textbf{Summary:}
\name takes advantage of the fine-grained elasticity of serverless to scale the aggregation service based on load changes, 
saving CPU consumption compared to serverful alternatives.
When comparing \name with \textbt{SL}, \name is even more compelling, with far lower CPU consumption because of \name's orchestration scheme and lightweight data plane (as we saw from the microbenchmark analysis). Thus, \name shows that it truly leverages the elasticity promise of the serverless computing paradigm. 
\hide{The optimized serverless infrastructure in \name also contributes to \name's cost savings, as demonstrated in the microbenchmark analysis.}

\vspace{-3mm}\section{Related work}\label{sec:related}\vspace{-2mm}
We have discussed the pros and cons of prior work on serverful~\cite{google-fl-mlsys19, papaya-mlsys2022} and serverless~\cite{adafed, lambda-fl, fedkeeper} FL systems in \S\ref{sec:background}. \name goes beyond these prior designs with an optimized serverless infrastructure and efficient orchestration to truly realize the promise of serverless computing. We now discuss work related to \name from other perspectives.

\noindent\textbf{Federated Learning:}
As a fast-evolving ML technology, a large body of work has been proposed for FL; the proposals in~\cite{fedavg, fedprox, fedbuff, qffl, google_adaptive_fl} focus on FL algorithms while others investigate how to select FL clients or datasets more intelligently~\cite{oort, auxo, refl, pisces-socc22, fedcs, fedbalancer, hybrid_fl, p2pfl, fedlesscan}. \cite{liu2023venn} seeks to schedule FL jobs across a shared set of FL clients with less contention and reduce job scheduling delays. These efforts are orthogonal to \name because \name focuses on system-level optimization of model aggregation of FL. This makes \name a good complement to these efforts by providing an efficient and high-performance FL system to bring various FL approaches to the ground.

Several open-source FL platforms, \eg Flame~\cite{flame-cisco}, FATE~\cite{fate}, OpenFL~\cite{openfl}, FedML~\cite{fedml}, IBM federated learning~\cite{ibm-fl} have been launched to facilitate the promotion and adoption of FL in both research and applications. These platforms assume themselves to be a serverful design with static, inflexible deployment, which makes them unprepared for large-scale FL. \name can be used as a representative case to guide the future development of these platforms.

\noindent\textbf{Serverless computing optimization:}
Recent advances in serverless computing have triggered extensive research endeavors dedicated to optimizing its system design. Significantly, a prominent amount of investigation revolves around the enhancement of resource provisioning, function deployment, load balancing~\cite{atoll, mu-socc21, sequoia, kraken, graf, hermod, ditto}, runtime overhead reduction~\cite{firecracker, sand-atc, faasm, sledge, sock-atc}, and mitigation of function startup delay~\cite{silvery-hotedge20, shahrad-atc20, lin2019mitigating, faascache, lukewarm, vhive, faasnet} within serverless platforms. Furthermore, substantial efforts have been directed towards addressing the data plane overheads inherent in serverless architectures~\cite{spright-sigcomm22, nightcore-asplos, faasm, pheromone-nsdi}, characterized by heavyweight function chaining and sidecar proxy. 

Our work, combines the advantages of data plane optimization (\ie shared memory for hierarchical aggregation, in-place message queuing, event-driven sidecars, etc), to unlock the full potential of serverless computing, facilitating efficient and cost-effective FL in the cloud.

\vspace{-3mm}\section{Conclusion}\vspace{-2mm}
\name is an optimized serverless FL system aimed at making FL more efficient and significantly lowering its operational cost.
\name adopts hierarchical aggregation to support FL at scale.
Its serverless infrastructure leverages shared memory processing to offer high-speed yet efficient intra-node data plane and event-driven sidecar functionality to facilitate communication within hierarchical aggregation. 
\name's orchestration scheme adjusts the aggregation hierarchy based on load and, maximizes the utilization of shared memory through intelligent placement and reuse of aggregation function instances, thus saving the cost.
Our evaluation shows that \name's optimized data and control planes improve the resource efficiency of the aggregation service by more than \textbf{5\X}, compared to existing serverless FL systems, with \textbf{2.7\X} reduction on time-to-accuracy for ResNet-18. 
\name also achieves \textbf{1.8\X} better efficiency and \textbf{1.6\X} speedup on time-to-accuracy than a serverful system.
In training ResNet-152 to reach 70\% accuracy, \name is \textbf{1.68\X} faster than an existing serverless FL system, while reducing CPU costs by \textbf{4.23\X}.


\vspace{-3mm}\section*{Acknowledgments}\vspace{-2mm}

We thank the US NSF for their generous support through grants CRI-1823270, CNS-1818971, and Cisco for their generous gift and support.

\bibliography{reference}
\bibliographystyle{mlsys2023}


\appendix

    \begin{figure}[b]
    \centering
        \includegraphics[width=\columnwidth]{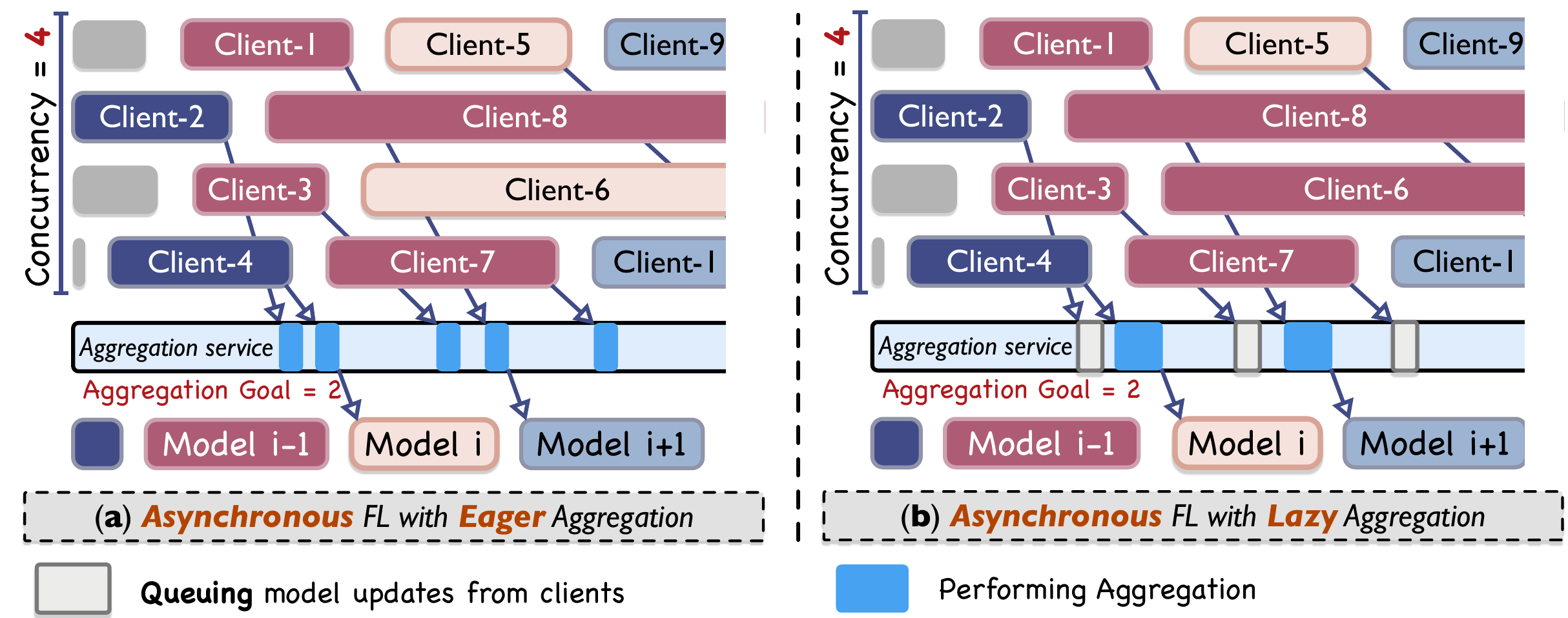}\vspace{-7mm}
    \caption{Asynchronous FL~\cite{papaya-mlsys2022} with different aggregation timing (``Eager'' and ``Lazy'').} 
    \label{fig:async-fl}
    \end{figure}

\hide{\section{FL-specific hyperparameters}\label{sec:fl-hyperpara}
There are several FL-specific hyperparameters that affect the behavior of the FL process, including {\it concurrency} and {\it aggregation goal}. {\it Concurrency} (Fig.~\ref{fig:sync-async-fl}), in both synchronous and asynchronous FL, indicates the number of participating clients, performing local training in parallel. A certain number of clients are selected to satisfy {\it concurrency}.
The {\it aggregation goal} specifies the expected number of model updates to be received before the global model is updated to the new version. This determines the timing for the model to be updated to the new version.

In Synchronous FL, over-provisioning the number of clients (over-selection) is often used to mitigate the impact of stragglers, specifying an {\it aggregation goal} to be less than the training concurrency. This leads to wasted effort by slow clients, whose updates may arrive late, thus discouraging participation by slow clients~\cite{refl}. Their speed may often be negatively correlated with the data quality~\cite{pisces-socc22}, ultimately affecting the model quality of Synchronous FL.

Asynchronous FL also requires that the {\it aggregation goal} be smaller than the concurrency, to avoid being impeded by potential stragglers. However, since Asynchronous FL allows stale updates to still be aggregated, {\it no} client effort will be wasted compared to the Synchronous FL. This ensures high client utilization throughout the FL process and potentially a more diverse dataset contributed by slow clients.}

\section{Message flow of intra-node and inter-node routing}\label{sec:routing-details}
    \begin{figure}[t]
    \centering
        \includegraphics[width=\columnwidth]{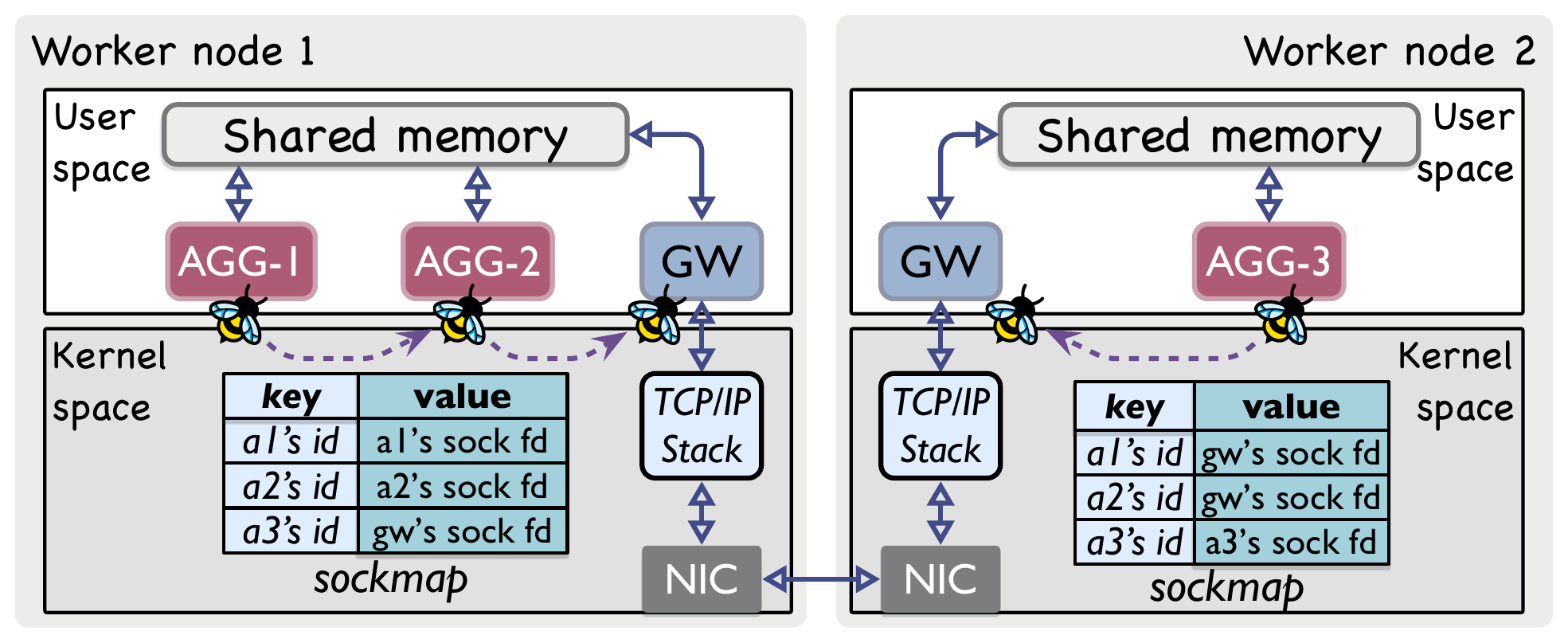}\vspace{-4mm}
    \caption{Intra-/inter-node direct routing within hierarchical aggregation.}
    \label{fig:direct-routing}
    \vspace{-6mm}
    \end{figure}

\noindent\textbf{Intra-node routing:}
\name makes full use of its shared memory support to facilitate {\it zero-copy} exchange of model updates between aggregators.
The shared memory object in \name is addressed by the object key, which is a 16 byte string randomly generated by the shared memory manager when it initializes shared memory objects. We also assign each aggregator a unique ID. The zero-copy data exchange between aggregators depends on delivering the object key, as the data is kept in place in shared memory.

\name utilizes eBPF's \skmsg (integrated in the eBPF-based sidecar), combined with eBPF's sockmap~\cite{Red2021Understanding}, to pass the object key between aggregators on the same node.
Upon receiving the object key, 
the \skmsg program uses the ID of the source aggregator as the key to look up the \textit{sockmap} to find the socket interface of the destination aggregator so that the object key may be delivered to it for access of the shared memory object.

\noindent\textbf{Inter-node routing:} When the source aggregator communicates with a destination aggregator on a different node, it sends the object key to the local gateway first. The local gateway uses the object key to retrieve the model update from shared memory and performs the necessary payload transformation. It then uses the source aggregator ID to look up the inter-node routing table to obtain the destination aggregator ID and the IP address of the remote node hosting the destination aggregator. The model update is sent through the remote node's gateway to the destination aggregator.
The remote gateway stores the received model update in shared memory and uses \skmsg to notify the destination aggregator, along with the local object key.

\noindent\textbf{Online hierarchy update:}
\name re-configures intra-/inter-node routes
each time the hierarchy is updated. 
The routing manager in the \name agent takes the DAG input (generated by the TAG, \S\ref{appendix:abstraction}) from the control plane that describes the connectivity between aggregators, and correspondingly updates routes into the inter-node routing table in the gateway and in-kernel
{\it sockmap},
using the userspace eBPF helper, \texttt{bpf\_map\_update\_elem()}~\cite{ebpf-helpers}.
The TAG describes the cross-level data dependency between aggregators.

\section{Model checkpoints}\label{appendix:model-checkpoints}
We support model checkpoints, where the model parameters are periodically saved to an external storage service to ensure data persistence and potential recovery in case of failures. The checkpointing occurs after the aggregator completes the aggregation of specified model updates, where the aggregator submits a request to the \name agent to perform model checkpoints asynchronously in the background. 
This prevents checkpoint delays from being added to the aggregation completion time.

\section{Message queueing flow in \name: Receive (RX) and Transmit (TX)}\label{appendix:msg-queueing}
On the receive (RX) path, protocol processing by the kernel TCP/IP stack is first performed. The gateway running in userspace receives the raw L7 payload from the kernel and then 
extracts the model updates (encoded as \texttt{tensor} data type), depending on the adopted L7 protocol (\eg gRPC, MQTT).
We convert the model update from \texttt{tensor} data type to \texttt{NumpyArray} before writing it to shared memory, as Python's {\tt multiprocessing} module 
does not support manipulation of the \texttt{tensor} data type.
On the transmit (TX) path, the reverse payload processing is done.

\section{Abstraction for fine-grained control}\label{appendix:abstraction}
To facilitate fine-grained control of \name's orchestration,
we treat an aggregator process within a sandboxed runtime (\eg container) as the atomic unit for management. 
The control plane needs a generic means to describe connectivity between components and placement affinity.
We make use of Topology Abstraction Graph (\textit{TAG}) in \flame~\cite{flame2023} to describe the aggregator-to-aggregator connectivity and aggregator-client connectivity.
Each node in such a graph is associated with a ``\textit{role}'' metadata, denoted as either aggregator or client. A ``\textit{channel}'' metadata denotes the underlying communication mechanism (\eg intra-node shared memory, inter-node kernel networking) used for  connectivity.

We configure the placement-affinity to facilitate locality-aware placement through the \textit{groupBy} attribute in the channel abstraction, which 
accepts a string as a label to specify a group. Therefore, keeping the same label in the attribute allows us to cluster roles into a group. 
The \name coordinator enables necessary orchestration decisions, \eg runtime reuse and locality-aware placement, through manipulation of these abstractions (role and channel).

\begin{figure}[b]
    \vspace{-4mm}
  \centering
        \subfigure[CPU cost]{
            \includegraphics[width=.3\columnwidth]{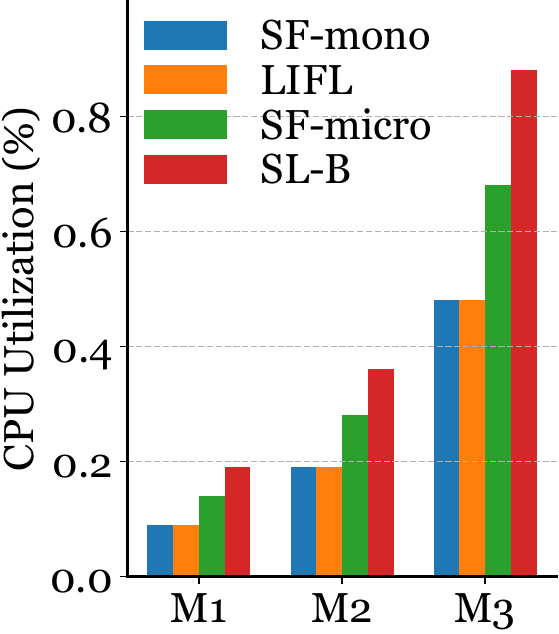}
            \label{fig:mq-cpu}
        } \hfill
        \subfigure[Memory cost]{
            \includegraphics[width=.3\columnwidth]{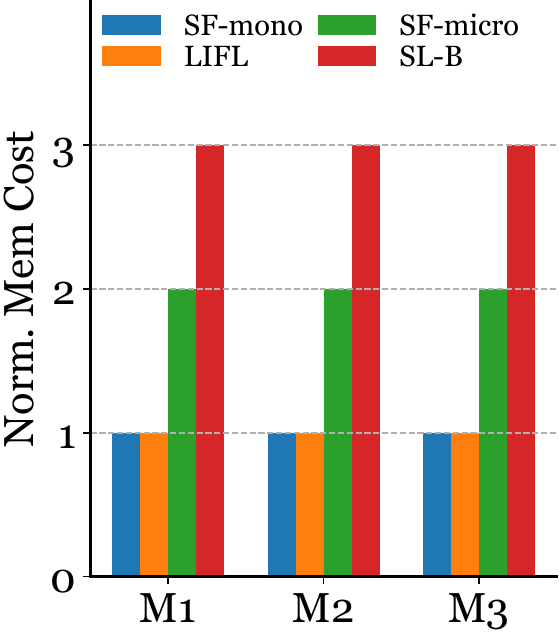}
            \label{fig:mq-mem}
        } \hfill
        \subfigure[End-to-end delay]{
            \includegraphics[width=.3\columnwidth]{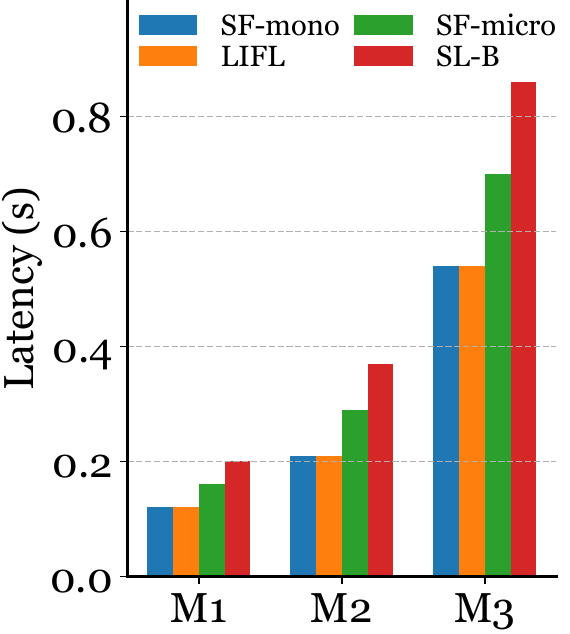}
            \label{fig:mq-latency}
        } 
  \caption{Message queuing overheads.}
  \label{fig:msg-queuing-overhead}
\end{figure}

\section{Maximum Service Capacity of Worker Nodes}\label{sec:mc}\vspace{-2mm}
\name actively monitors both $E_{i,t}$
and the arrival rate $k_{i,t}$ using the sidecar in \S\ref{sec:ebpf-sidecar}.
We determine the value of $MC_{i}$ offline. We incrementally increase the arrival rate $k_{i}$ to node $i$. Let $k'_i$ and $E'_t$ denote the arrival rate and average execution time at the point we observe a significant increase in $E_{i}$. This indicates that node $i$ is becoming overloaded and we estimate $MC_{i}$ as $k'_{i} \times E'_{i}$.

\section{In-place message queueing benefit}\label{sec:msg-queue-expt}\vspace{-2mm}
We examine \name's in-place message queuing through a comparison with the serverful and serverless alternatives depicted in Fig.~\ref{fig:mq-alternatives}, including the {\it monolithic} serverful setup (denoted as \textbt{SF-mono}), the {\it microservice}-based serverful setup (denoted as \textbt{SF-micro}), and the {\it basic} serverless setup (denoted as \textbt{SL-B}).
We quantify the overheads of message queuing for a single model update transfer between
the client to the aggregator.
We consider three metrics: (1) the total memory consumed for
queuing the model update along the data pipeline; (2) the CPU cycles spent in the data pipeline;
and (3) the end-to-end networking delay from the client to the aggregator.
Note that we exclude the overhead on the client-side.
We consider three ML models with distinct sizes: (\textbt{M1}) ResNet-18 ($\sim$44MB), (\textbt{M2}) ResNet-34 ($\sim$83MB), and (\textbt{M3}) ResNet-152 ($\sim$232MB).

Fig.~\ref{fig:msg-queuing-overhead} shows the results of CPU, memory cost and end-to-end networking delay. The memory consumption in \textbt{SF-mono} is mainly from the in-memory queue inside the aggregator. For \name it is primarily consumed by the shared memory used to buffer the model update.
But, \textbt{SL-B} consumes 3\X more memory than \textbt{SF-mono} and \name. The extra memory consumption of \textbt{SL-B} comes from the use of sidecar and message broker, both of which need to locally buffer the model update. \textbt{SF-micro}, on the other hand, saves one queuing stage at the sidecar, but still incurs the queuing at the message broker and consuming extra memory.
\name's in-place message queuing totally eliminates these unnecessary queuing stages.

Looking at the CPU consumption, \name is $\sim$1.5\X and $\sim$1.9\X less than \textbt{SL-B} and \textbt{SF-micro}, respectively. In terms of the end-to-end networking delay (client to aggregator), \name is $\sim$1.3\X and $\sim$1.7\X less than \textbt{SL-B} and \textbt{SF-micro}, respectively. \name's improvement in CPU cost and networking delay, compared to \textbt{SL-B} and \textbt{SF-micro}, are also a result of the elimination of the sidecar and message broker from the data pipeline, and the message queuing if far more efficient. 
This illustrates the benefits of \name's in-place message queuing, achieving the equivalent efficiency and performance of a monolithic, serverful design (with far less resource consumption as we see for typical FL applications).

\hide{We additionally evaluate the inter-node dataplane, where the leaf aggregators are placed on a different node than the top aggregator.
As shown in Fig.~\ref{fig:p2p-latency-inter} and~\ref{fig:p2p-cpu-inter}, for different message sizes,
\textbt{SL} consistently achieves {\color{red} XX\%} higher latency than \textbt{SF} and \name. At the same time, it uses {\color{red} XX\% to XX\%} more CPU. 
This clearly shows the poor efficiency and performance of the indirect networking used in serverless setup, \new{caused by its use of the message broker and heavyweight sidecar.} \cut{We further quantify the CPU and latency specifically incurred by the message broker and sidecar in \textbt{SL}. The result shows that the message broker accounts for {\color{red} XX\%} of the total overhead (both CPU and latency), and the sidecar accounts for {\color{red} XX\%} of the total overhead, which is also quite significant.}
Although \name involves the gateway to support inter-node communication between aggregators, the communication between the local aggregator and the gateway is through shared memory (Fig.~\ref{fig:direct-routing}), incurring negligible overhead. The major dataplane overhead comes from the communication between gateways on the different worker nodes, which is equal to that of the direct networking between aggregators using the \textbt{SF} framework.
The result is that the inter-node dataplane performance of \name is quite close to \textbt{SF}.
Further, through careful placement, aggregators in \name can benefit from high-speed intra-node dataplane over shared memory, as discussed next.}

\subsection{Stateful ``tax'' in \name}\vspace{-2mm}
The per-node gateway is a key component that enables a number of data plane functionalities in \name, including in-place message queuing and inter-node data transfer. Unlike stateless aggregators, the gateway is deployed as a stateful, persistent component on every \name worker node. This raises the concern about the stateful ``tax'', \ie the CPU/memory cost of having stateful components in the FL system. 

On the other hand, a stateful ``tax'' of some form commonly exists in serverful and serverless alternatives, as shown in Fig.~\ref{fig:mq-alternatives}. The stateful component in a monolithic serverful setup is the aggregator itself, running as an ``always-on'' monolith. 
In the microservice-based serverful setup, the message broker is the stateful component, as is the case for the basic serverless setup.
We quantitatively compare the stateful ``tax'' of \name's gateway with serverful and serverless alternatives in Fig.~\ref{fig:mq-alternatives}. The result in \S\ref{sec:msg-queue} shows that stateful ``tax'' in \name is the lowest.




    \begin{figure}[htbp]
    \centering
        \includegraphics[width=.8\columnwidth]{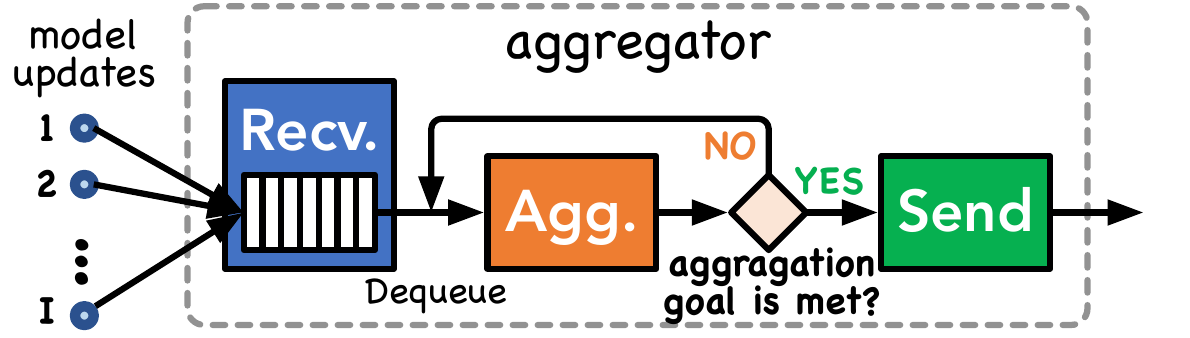}\vspace{-4mm}
    \caption{Step-based processing model.}
    \label{fig:step-model}
    \end{figure}

\section{Step-based Processing Model}\label{sec:step-model}
The basic processing model of an \name aggregator can be abstracted as a multiple-producer, single-consumer pattern, as shown in Fig.~\ref{fig:step-model}. Multiple upstream producers (clients or aggregators) are mapped to a single consumer (aggregator only). The single consumer gathers model updates from assigned producers and computes the aggregated model update.


Looking deeper into the aggregator, \name adopts a step-based processing model. At the core of this design is a processing pipeline of three steps: {\bf (1)} \texttt{Recv:} Receive model updates from all assigned producers. The received model update is enqueued in a FIFO queue. In \name, the object key of the model update is enqueued as the actual model update resides in shared memory; {\bf (2)} \texttt{Agg:} Aggregator dequeues a model update from the FIFO queue in \texttt{Recv} and then aggregates it. The \texttt{Agg} step checks if the aggregation goal is met after the dequeued update is aggregated. 
If the aggregation goal is not met, \texttt{Agg} is repeated until the aggregation goal is met, before moving to \texttt{Send}; and {\bf (3)} \texttt{Send:} sends the final model update to the designated consumer.
The execution of \texttt{Recv} and \texttt{Agg} overlaps to enable eager aggregation, \ie once the \texttt{Recv} step receives a model update, it immediately passes the model update to \texttt{Agg} step for aggregation.

\hide{Our step-based processing model supports both lazy aggregation and eager aggregation (\S\ref{sec:sync-async-fl}). In the lazy mode, the \texttt{Agg} step waits for the completion of \texttt{Recv} before starting. 
In the eager mode, the execution of \texttt{Recv} and \texttt{Agg} can overlap, \ie once the \texttt{Recv} step receives a model update, it immediately passes the model update to \texttt{Agg} step for aggregation. 
In both lazy and eager modes, \texttt{Send} needs to wait for \texttt{Agg} to finish (\ie when the aggregation goal is achieved). The choice of lazy or eager mode depends on the requirements of the FL algorithm, as described in \S\ref{sec:sync-async-fl}.}

\end{document}